\documentclass[10pt,final,journal,letterpaper]{IEEEtran}
\usepackage{color,graphicx,psfrag,epsfig,epsf,latexsym,theorem,hhline,amsmath,amssymb,array}
\usepackage{setspace}
\IEEEoverridecommandlockouts
\usepackage{cite}
\usepackage{amsmath,amssymb,amsfonts}
\usepackage{algorithmic}
\usepackage{graphicx}
\usepackage{textcomp}
\def\BibTeX{{\rm B\kern-.05em{\sc i\kern-.025em b}\kern-.08em
    T\kern-.1667em\lower.7ex\hbox{E}\kern-.125emX}}
\usepackage[ruled]{algorithm2e}

\usepackage{cite}

\newcommand {\mymarginpar}[1]{\marginpar{#1}}
\renewcommand {\marginpar}[1]{}

\def\_{\rule{.3em}{.15ex}}      % Get underscore by typing \_.

\newcommand{\ls}[1]
   {\dimen0=\fontdimen6\the\font
    \lineskip=#1\dimen0
    \advance\lineskip.5\fontdimen5\the\font
    \advance\lineskip-\dimen0
    \lineskiplimit=.9\lineskip
    \baselineskip=\lineskip
    \advance\baselineskip\dimen0
    \normallineskip\lineskip
    \normallineskiplimit\lineskiplimit
    \normalbaselineskip\baselineskip
    \ignorespaces
   }
\newcommand {\bearn}{\begin{eqnarray*}}
\newcommand {\eearn}{\end{eqnarray*}}
\newcommand {\barr}{\begin{array}}
\newcommand {\earr}{\end{array}}

\newcommand {\N}{{\cal N}}

\newtheorem{definition}{Definition}
\newtheorem{property}[definition]{Property}
\newtheorem{proposition}[definition]{Proposition}
\newtheorem{lemma}[definition]{Lemma}
\newtheorem{theorem}[definition]{Theorem}
\newtheorem{corollary}[definition]{Corollary}
\newtheorem{example}[definition]{Example}
\newtheorem{remark}[definition]{Remark}

\newcommand{\comb}[2]
{\left ( \begin{array}{c} #1 \\#2 \end{array} \right ) }

\newcommand {\benum} {\begin{enumerate}}
\newcommand {\eenum} {\end{enumerate}}

\newcommand {\bdesc} {\begin{description}}
\newcommand {\edesc} {\end{description}}

\newcommand {\bfig}[2] {\begin{figure}
  \centering
  \includegraphics[width=#2]{#1}}
\newcommand {\brotatefig}[2] {\begin{figure}[htbp]
                        \centerline {
                         \epsfig{figure={#1},clip=,angle=-90,width={#2}}}}
\newcommand {\bfigfirst}[2] {\begin{figure}[h]
                        \centerline {
                        \setlength{\epsfxsize}{#2}
                        \epsffile{#1}}}
\newcommand {\efig}[2]{ \caption{#2}
                        \label{fig:#1}
                        \end{figure}
                        \mymarginpar{fig:#1}}
\newcommand {\erotatefig}[2]{ \caption{#2}
                        \label{fig:#1}
                        \end{figure}
                        \mymarginpar{fig:#1}}
\newcommand {\rfig}[1]{Figure \ref{fig:#1}}

\newcommand {\btab}[1]{
                       \begin{table}
                       \centering
                       \begin{tabular}{#1}}
\newcommand {\etab}[3] {
                       \end{tabular}
                       \caption[#3]{#2}
                       \label{tab:#1}
                       \end{table}
                       \mymarginpar{tab:#1}
                       \vspace{.1in}}

\newcommand {\btabular}[1]{\begin{center}
                       \begin{tabular}{#1}}
\newcommand {\etabular}{\end{tabular}
                       \end{center}}

\newcommand {\bdefin}[1]{\begin{definition}
                      \mymarginpar{def:#1}
                      \label{def:#1} }
\newcommand {\edefin}       {\end{definition}}
\newcommand {\rdef}[1]{Definition \ref{def:#1}}

\newcommand {\bpro}[1]{\begin{property}
                      \mymarginpar{pro:#1}
                      \label{pro:#1} }
\newcommand {\epro}   {\end{property}}

\newcommand {\bprop}[1]{\begin{proposition}
                      \mymarginpar{prop:#1}
                      \label{prop:#1} }
\newcommand {\eprop}       {\end{proposition}}
\newcommand {\rprop}[1]{Proposition \ref{prop:#1}}

\newcommand {\blem}[1]{\begin{lemma}
                      \mymarginpar{lem:#1}
                      \label{lem:#1} }
\newcommand {\elem}   {\end{lemma}}
\newcommand {\rlem}[1]{Lemma \ref{lem:#1}}

\newcommand {\bthe}[1]{\begin{theorem}
                      \mymarginpar{the:#1}
                      \label{the:#1} }
\newcommand {\ethe}   {\end{theorem}}
\newcommand {\rthe}[1]{Theorem \ref{the:#1}}

\newcommand {\bproof}{\noindent {\bf Proof.} \ }
\newcommand {\eproof} {\hfill \squares \\ \vspace{.3cm}}
\newcommand {\bcor}[1]{\begin{corollary}
                      \mymarginpar{cor:#1}
                      \label{cor:#1} }
\newcommand {\ecor}   {\end{corollary}}
\newcommand {\rcor}[1]{Corollary \ref{cor:#1}}

\newcommand {\bax}[1]{\begin{axiom}
                      \mymarginpar{ax:#1}
                      \label{ax:#1} }
\newcommand {\eax}       {\vspace{-.1in} \end{axiom}}

\newcommand {\bex}[2]{\vspace{.1in}
                      \begin{example}
                      \mymarginpar{ex:#1}
                       {\bf #2}
                      \label{ex:#1} }
\newcommand {\eex}       {\end{example} \vspace{.3cm} }

\newcommand {\brem}[1]{\begin{remark}
                      \mymarginpar{rem:#1}
                      \label{rem:#1} \em }
\newcommand {\erem}   {\end{remark}}

\newcommand {\beq}[1]{\mymarginpar{eq:#1}
                      \begin{equation}
                      \label{eq:#1} }

\newcommand {\beqno}[1]{\mymarginpar{eq:#1}
                      \begin{eqnarray}
                      \nonumber}

\newcommand {\eeq}       {\end{equation}}
\newcommand {\eeqno}       { && \end{eqnarray}}
\newcommand {\req}[1]{(\ref{eq:#1})}

\newcommand {\bear}[1]{\mymarginpar{eq:#1}
                       \begin{eqnarray}
                       \label{eq:#1} }

\newcommand {\bearno}[1]{\mymarginpar{eq:#1}
                       \begin{eqnarray}
                       \nonumber}

\newcommand {\eear}{\end{eqnarray}}
\newcommand {\eearno}{\end{eqnarray}}
\newcommand {\bsel}{\left \{ \begin{array}{cl}}
\newcommand {\esel}{\end{array} \right.}

\newcommand {\bmat}[1]{\left [ \begin{array}{#1}}
\newcommand {\emat}{\end{array} \right ]}
\newcommand {\bsec}[2]{\mymarginpar{sec:#2}
                       \section{#1}
                       \label{sec:#2} }

\newcommand {\rsec}[1]{Section \ref{sec:#1}}

\newcommand {\bsubsec}[2]{\mymarginpar{sec:#2}
                       \subsection{#1}
                       \label{sec:#2} }

\def\R{I\kern-0.30em R}
\def\N{I\kern-0.30em N}
\def\P{I\kern-0.30em P}
\newcommand\squares{\vrule height6pt width7pt depth1pt}

\def\ex{{\bf\sf E}}
\def\pr{{\bf\sf P}}

\newcommand{\betax}{X}
\newcommand{\betab}{\beta}
\newcommand{\alphax}{X}

\begin{document}

\title{{\bf A Quasi-random Algorithm for Anonymous Rendezvous in Heterogeneous Cognitive Radio Networks}}
%\title{{\bf IMC: An Interleaved Modular Clock Algorithm for Asynchronous Multiuser Rendezvous with Unique User %IDs}}
%\title{{\bf To Spread Out or Stick Together: a Strategic Question for Asynchronous Multiuser Rendezvous}}
% in Heterogeneous Cognitive Radio Networks}}

%\iffalse
\author{Cheng-Shang Chang, Yeh-Cheng Chang and Jang-Ping Sheu\\
Department of Computer Science and Institute of Communications Engineering\\
National Tsing Hua University \\
Hsinchu 30013, Taiwan, R.O.C. \\
Email:  cschang@ee.nthu.edu.tw; jas1123kimo@gmail.com;  sheujp@cs.nthu.edu.tw}
%\thanks{The conference version of this paper \cite{ICC} was accepted for presentation in ICC 2018.}}
%\fi

\maketitle

\begin{abstract}
The multichannel rendezvous problem that asks two secondary users to rendezvous on a common available channel in a  cognitive radio network (CRN) has received a lot of attention lately. Most rendezvous algorithms in the literature focused on constructing channel hopping (CH) sequences that guarantee finite maximum time-to-rendezvous (MTTR). However, these algorithms perform rather poorly in terms of the expected time-to-rendezvous (ETTR) even when compared to the simple random algorithm.
%One of the long lasting open questions for the multichannel rendezvous problem is whether it is possible to have a rendezvous algorithm that has a comparable ETTR to the random algorithm and a comparable MTTR to the best bound in the literature. To address such a question,
In this paper, we propose the quasi-random (QR) CH algorithm  that has a comparable ETTR to the random algorithm and a comparable MTTR to the best bound in the literature. Our QR algorithm does not require the unique identifier (ID) assumption and it is very simple to implement in the  symmetric,  asynchronous, and heterogeneous setting with multiple radios. In a CRN with $N$ commonly labelled channels, the  MTTR of the QR algorithm is bounded above by $9 M \lceil n_1/m_1 \rceil \cdot \lceil n_2/m_2 \rceil$ time slots, where $n_1$ (resp. $n_2$) is the number of available channels to user $1$ (resp. 2), $m_1$ (resp. $m_2$) is the number of radios for user $1$ (resp. 2), and $M=\lceil \lceil \log_2 N \rceil /4 \rceil *5+6$. Such a bound is only slightly larger than the best $O((\log \log N) \frac{n_1 n_2}{m_1 m_2})$ bound in the literature. When each SU has a single radio, the ETTR is bounded above by $\frac{n_1 n_2}{G}+9Mn_1n_2 \cdot (1-\frac{G}{n_1 n_2})^M$, where  $G$ is the number of common channels between these two users.
By conducting extensive simulations, we show that for both the MTTR and the ETTR, our algorithm is comparable to the simple random algorithm and it outperforms several existing algorithms in the literature.

\end{abstract}


%\bsec{Introduction}{introduction}

%\IEEEPARstart{W}{ireless} networks

%\bsec{Problem statement}{statement}
%\bsec{Introduction}{introduction}

%Motivated by setting up a common control channel in a cognitive radio network
%In the multichannel rendezvous problem \cite{Bian2013,MOR2014}, there are
\bsec{Introduction}{introduction}

%\IEEEPARstart{I}{n}
In a cognitive radio network (CRN), there are a set of frequency channels that are shared by  two types of spectrum users: primary users (PUs) and secondary users (SUs). PUs have dedicated channels  assigned to them. On the other hand, SUs can only access channels that are not being used by PUs. As such,
 SUs need to  sense a number of frequency channels that are not used by PUs. Such a set of channels is called the {\em available channel set} for an SU. In order for two SUs to communicate with each other, they need to find a common available channel. Such a problem is known as the multichannel rendezvous problem in a CRN and it is usually solved in a {\em distributed} manner by hopping over the available channels over time. For the multichannel rendezvous problem, it is thus important to design channel hopping (CH) sequences so as to minimize the time-to-rendezvous (TTR).

%\cite{SSCH,Seq2008,Walrand2008,SynMAC,Quorum,Shih10,CRSEQ,DRSEQ,Theis2011,ETCH,JS,Hou2011,Coor2012,Bian13,Lin13,DRDS13,ChangGY13,Gu14,Obli2014,Chen14,Yu14,Chuang14,Multiradio14,ARCH,CBH2014,LChen14,Krunz2015,MOR2014,Gu15,LChen15,Efficient15,ToN2015,Li15,Yaday15}).
For the multichannel rendezvous problem, there are many CH schemes proposed in the literature
  (see e.g.,
\cite{CRSEQ,Theis2011,JS,Bian13,DRDS13,Gu14,Chen14,Yu14,Chuang14,Multiradio14,ARCH,CBH2014,LChen14,Gu15,LChen15,Krunz2015,Li15,Yaday15,Chang17,Zhao18,RPS,AMRR,ICC}).  As discussed in the tutorial \cite{Tutorial} and the book \cite{Book}, CH schemes can be classified into various categories depending on their assumptions.
%\begin{enumerate}
% \item

\noindent 1) {\bf Asymmetric vs. symmetric}: In a symmetric CH scheme, users follow the same strategy to generate their CH sequences. On the other hand,  asymmetric algorithms (see e.g.,  \cite{Li15}, \cite{Yaday15},  \cite{Yu14}, \cite{Zhao18}) can assign users  different roles so that they can follow different strategies to generate their CH sequences. For instance, a user can be assigned the role of a sender or receiver.  The receiver can stay on the same channel while the sender cycles through all the available channels.  Since users follow different strategies, the time-to-rendezvous can be greatly reduced by using  asymmetric algorithms.

%\item
\noindent 2) {\bf Onymous vs. anonymous}: One simple way to assign different roles to users is by their identifiers (ID). In \cite{Li15}, \cite{Bian13}, \cite{LChen14}, \cite{Chuang14}, \cite{Gu14}, \cite{Chang17}, \cite{Zhao18}, it is assumed that each user is assigned with a unique ID, e.g., a MAC address. As such, users can map their IDs to play different roles to speed up the rendezvous process.

%     \item
\noindent 3) {\bf Synchronous vs. asynchronous}: A CH scheme is {\em synchronous} if the clocks (i.e., the indices of
time slots) of both SUs are the same. Synchronous CH schemes can achieve better performance than asynchronous CH schemes as both SUs can start their CH sequences simultaneously. However, in a distributed environment it might not be practical to assume that the clocks of two users are synchronized as they have not rendezvoused yet. Without clock synchronization, guaranteed rendezvous is much more difficult. In the literature, there are various asynchronous algorithms  (see e.g.,    \cite{Li15}, \cite{Yaday15}, \cite{Yu14}, \cite{Bian13}, \cite{LChen14}, \cite{Chuang14}, \cite{Gu14}, \cite{JS}, \cite{Chen14}, \cite{LChen15}, \cite{Gu15}).

%\item
\noindent 4) {\bf Homogeneous vs. heterogeneous}: A CH scheme is called {\em homogeneous} if the available channel sets of the two SUs are the same. On the other hand, it is called {\em heterogeneous} if the available channel sets of the two SUs are different. Two SUs that are close to each other are likely to have the same available channel sets. Due to the limitation of the coverage area of a user, two SUs tend to have different available channel sets if they are far apart. Rendezvous in a homogeneous environment is in general much easier than that in a heterogeneous environment.
There are various heterogeneous CH algorithms that have bounded TTR (see e.g.,   \cite{Li15}, \cite{Yaday15}, \cite{JS}, \cite{Chen14}, \cite{Gu15}). We note that in the literature some authors refer a homogenous (resp. heterogeneous) environment as a symmetric (resp. asymmetric) environment.

%\item
\noindent 5) {\bf Oblivious vs. non-oblivious}:  In most previous works for the multichannel rendezvous problem, it is commonly assumed that there is a universal channel labelling. As such, it is possible for a user to learn from a failed attempt to rendezvous. On the other hand,
oblivious rendezvous (see e.g., \cite{Multiradio14}, \cite{AMRR},  \cite{LChen14}, \cite{Chang17}) is referred to as the setting where nothing can be learned from  a failed attempt to rendezvous.

%\item
\noindent 6) {\bf Single radio vs. multiple radios}: Recently, several research works focus on the multi-radio CH schemes \cite{Multiradio14}, \cite{RPS}, \cite{AMRR}, \cite{ICC}. SUs equipped with multiple radios can generate CH sequences that hop on more than one channel in a time slot. This improves the probability of rendezvous and thus shortens the time-to-rendezvous.
%\end{enumerate}

As pointed out in the recent paper \cite{Chang17}, most works in the literature focused on deriving bounds for maximum time-to-rendezvous (MTTR), and
they perform rather poorly in terms of expected time-to-rendezvous (ETTR) even when compared to the simple random algorithm.
The rationale behind that is because there is usually a ``stay'' mode in these CH schemes. When an SU is  in its  ``stay'' mode, it stays on the same channel for a rather long period of time. As such, it is very likely that two SUs stay on two different channels for a long period of time.
To address the large ETTR problem, a hybrid CH algorithm was proposed in \cite{LChen15} for a homogeneous CRN. The idea is to interleave the simple random algorithm with a periodic CH algorithm that has a bounded MTTR, such as  CRSEQ \cite{CRSEQ} and JS \cite{JS}.
%Interleaving is done by a periodic wake-up sequence that chooses the random algorithm when it is in the asleep mode and the CH algorithm when it is in the awake mode. Such a hybrid CH algorithm can greatly reduce the ETTR by decreasing the duty cycle of the wake-up sequence (so as to increase the chance to be in the asleep mode).
However, the hybrid CH algorithm can only be used in a homogeneous CRN.

In  \cite{Chang17},
 the authors considered the {\em oblivious} rendezvous problem in heterogeneous CRNs and proposed a CH algorithm such that its ETTR is comparable to that of the random algorithm while its MTTR is still upper bounded by a finite constant.
 This is done by assuming there is a unique ID assigned to each user.
One of the problems of such an approach is that the length of an ID is usually very long, e.g.,  a MAC address contains 48 bits.
As the MTTR bound in \cite{Chang17} is proportional to the length of an ID, the MTTR bound could also be large in practice.
On the other hand, using the (mapped) ID to generate CH sequences makes it difficult for an SU to remain {\em anonymous}.
In particular, if the ID of a user is known to an adversary, then it could be used by the adversary to construct the same CH sequence for jamming attack \cite{Krunz2015}.
Thus, for the security reason it is crucial to eliminate the need of the unique ID assumption for each SU in \cite{Chang17}.

Without the unique ID assumption for each SU in \cite{Chang17},
the question is then whether it is still possible to have a  rendezvous algorithm  that has a comparable ETTR to the random algorithm and a comparable MTTR to the best bound in the literature.
Such a question is not only of theoretical interest but also of practical importance as the random algorithm outperforms most rendezvous algorithms in the literature regarding ETTR (despite its lack of theoretical guarantee for MTTR).
To address such a question, we extend the construction in \cite{Chang17} by proposing a quasi-random CH algorithm in this paper.
% that behaves as if were a random algorithm for every consecutive $M$ slots and
{\em The main idea of our quasi-random algorithm is to select at random an arbitrary channel in the available channel set of an SU as its ID (channel).
By doing so, we can leverage the construction in \cite{Chang17} that maps a binary ID to a CH sequence. The problem is that the unique ID assumption in \cite{Chang17} is no longer valid as the two SUs might select one of their common channels as their IDs.
To deal with such a problem, our second idea is to extend a binary ID to a ternary ID with elements in $\{0,1,2\}$.
When the symbol "2" appears, an SU simply stays on the channel that is used as its ID. By doing so, SUs with the same ID are still guaranteed to rendezvous.
}

%\bsec{The multichannel rendezvous problem}{formulation}

Our setting for the multichannel rendezvous problem is  the {\em symmetric, anonymous, asynchronous, and heterogeneous setting with multiple radios}.
However, we do assume that there is a universal channel labelling.
Specifically, we consider a CRN with $N$ channels (with $N \ge 2$), indexed from $0$ to $N-1$. Time is slotted (the discrete-time setting) and indexed from $t=0,1,2,\ldots$. There are two users who would like to rendezvous on a common available channel by hopping over these $N$ channels with respect to time.
 The available channel set for user $i$, $i=1,2$, is $${\bf c}_i=\{c_i(0), c_i(1), \ldots, c_i(n_i-1)\},$$ where $n_i=|{\bf c}_i|$ is the number of available channels to user $i$, $i=1,2$.  We assume that there is at least one channel that is commonly available to the two users, i.e.,
\beq{avail1111}
{\bf c}_1 \cap {\bf c}_2 \ne \phi.
\eeq
Moreover, we assume that user $i$ has
  $m_i$ radios, where $m_i \ge 1$, $i=1$ and 2.
  Denote by $X_1(t)$ (resp. $X_2(t)$) the set of channels  selected by user 1 (resp. user 2) on its $m_i$ radios at time $t$ (of the global clock).
  Then the time-to-rendezvous (TTR), denoted by  $T$, is the number of time slots (steps) needed for these two users to select a common available channel, i.e.,
\beq{meet1111}
T=\inf\{t \ge 0: X_1(t) \cap X_2(t)  \ne \phi\}+1,
\eeq
where we add 1 in \req{meet1111} as we start from $t=0$.

%Under such a setting, we propose a
For the quasi-random algorithm, we have the following theoretical results:
%We summarize our contributions as follows:

\noindent (i)  The  MTTR is bounded above by $9 M \lceil n_1/m_1 \rceil \cdot \lceil n_2/m_2 \rceil$ time slots, where  $M=\lceil \lceil \log_2 N \rceil /4 \rceil *5+6$. Such a bound is only slightly larger than the best $O((\log \log N) \frac{n_1 n_2}{m_1 m_2})$ bound in the literature (see e.g., \cite{ICC} and references therein).

\noindent (ii) When each SU has a single radio, the ETTR is bounded above by $\frac{n_1 n_2}{G}+9Mn_1n_2 \cdot (1-\frac{G}{n_1 n_2})^M$,
where  $G$ is the number of common channels between these two users. Note that the first term is the ETTR of the random algorithm and the second term approaches 0 when $M \to \infty$. Thus, the ETTR of the quasi-random algorithm is almost the same as that of the random algorithm when $M$ is large.

% Our algorithm is  backward compatible with users equipped with a single radio.

%\noindent (iii)
By conducting extensive simulations, we show that for both the MTTR and the ETTR, our algorithm is comparable to the simple random algorithm and it outperforms several existing algorithms, including
JS/I  \cite{RPS}, GCR \cite{Multiradio14}, RPS \cite{RPS},  AMRR \cite{AMRR} and FMRR \cite{ICC}.

The rest of this paper is organized as follows: In \rsec{two}, we consider the two-user rendezvous problem and show how one can construct the CH sequences from the quasi-random algorithm.  In \rsec{exp}, we  conduct extensive simulations to compare the performance of our quasi-random algorithm with that of some best-performed channel hopping algorithms in the literature.
Finally, we conclude the paper in \rsec{con}.

\bsec{Constructions of the CH sequences}{two}

As mentioned in \rsec{introduction}, our main idea is to leverage the construction of the CH sequences in \cite{Chang17} by selecting at random an arbitrary channel in the available channel set of an SU as its ID.
For this, in \rsec{strong} we first generalize the concept of the strong symmetrization mapping in \cite{Chang17} to map a ternary ID to a CH sequence. We show in \rsec{4B5B} that the 4B5B  encoding scheme can be used as a strong ternary symmetrization mapping. In \rsec{modular}, we then propose the quasi-random algorithm.

\bsubsec{Strong ternary symmetrization mapping}{strong}

We first generalize  the concept of strong symmetrization class in \cite{Chang17} for binary vectors to ternary vectors with the elements in
$\{0,1,2\}$. A ternary digit in $\{0,1,2\}$ is called a {\em trit} in this paper.

%Such a concept is stronger than the cyclic unique property for the symmetrization class in \cite{Group2015}.

\bdefin{strong}{\bf (Strong ternary symmetrization mapping)}
Consider a set of $M$-trit codewords (with size $K$)
$$\{{\bf w}_i=(w_i(0),w_i(1), \ldots, w_i(M-1)), i=1,2, \ldots, K\}.$$
Let
\bearn
&&\mbox{Rotate}({\bf w}_i,d)\\
&&=(w_i(d),w_i(d+1), \ldots, w_i((d+M-1)\;\mbox{mod}\;M)),
%\\
%&&=(\tilde w_i(0),\tilde w_i(1), \ldots, \tilde w_i(M-1)),
\eearn
be the vector obtained by cyclically shifting the vector ${\bf w}_i$ $d$ times.
%for $i=1,2, \ldots, K.$$
Then this set of codewords is called a {\em strong ternary $M$-symmetrization class} if
$w_i(0)=2$ for all $i$, and for either the time shift $(d\;\mbox{mod}\;M) \ne 0$ or $i \ne j$,  (at least) one of  the following two properties is satisfied:
\begin{description}
\item[(i)] There exist $0 \le \tau_1, \tau_2 \le M-1$ such that $w_i(\tau_1)=1, w_j((\tau_1+d)\;\mbox{mod}\;M)=0$ and
$w_i(\tau_2)=0, w_j((\tau_2+d)\;\mbox{mod}\;M)=1$.
\item[(ii)] There exist $0 \le \tau_1, \tau_2 \le M-1$ such that $w_i(\tau_1)=w_j((\tau_1+d)\;\mbox{mod}\;M)=1$,
 and $w_i(\tau_2)\ne w_j((\tau_2+d)\;\mbox{mod}\;M)$.
%\item[(iii)] There exist $0 \le \tau_1, \tau_2 \le M-1$ such that $w_i(\tau_1)=w_j((\tau_1+d)\;\mbox{mod}\;M)=1$,
% and $w_i(\tau_2)\ne w_j((\tau_2+d)\;\mbox{mod}\;M)$.
\end{description}
A one-to-one mapping from the set of  integers $[1,\ldots, K]$ to a strong ternary $M$-symmetrization class is called a {\em strong ternary $M$-symmetrization mapping}.
\edefin

In comparison with the original definition of the strong symmetrization class in \cite{Chang17}, here we require that the first trit of every vector is 2. Also, we replace the condition in (ii) by $w_i(\tau_1)=w_j((\tau_1+d)\;\mbox{mod}\;M)=1$ (instead of 0).
Also, we note that the strong ternary symmetrization mapping is stronger than the ``ternary symmetrization mapping'' in Lemma 2 of \cite{Chuang14} that only requires the codeword to be cyclically unique.
%As pointed out in \cite{Chang17}, s
Such a stronger property enables us to construct CH sequences that behave as if they were random.

%@@@ This should not affect anything from symmetry @@@
%Clearly, a strong $M$-symmetrization class is an $M$-symmetrization class in \cite{Group2015}.
%We note that if $i=j$ and $(d\;\mbox{mod}\;M) = 0$, then $w_i(\tau)=\tilde w_i(\tau)$ for all $\tau$.

\bsubsec{4B5B encoding}{4B5B}

%\iffalse
{\tiny
\begin{table}
\begin{center}
\caption{The 4B5B encoding table}{%
    \begin{tabular}{||c|c||c|c||}
        \hline
        4B data & 5B code & 4B data & 5B code \\ \hline
 0000  & 11110 & 1000 & 10010\\ \hline
 0001 & 01001 &1001 & 10011  \\ \hline
 0010  & 10100 & 1010& 10110\\ \hline
 0011 & 10101 & 1011 &10111\\ \hline
 0100 &　01010 &1100 &11010 \\ \hline
0101 & 01011 & 1101 &11011\\ \hline
0110 & 01110 &  1110 & 11100\\ \hline
 0111 &  01111 &1111 &11101 \\
       \hline
    \end{tabular}}
        \label{table:4B5B}
\end{center}
\end{table}
}
%\fi

Analogous to \cite{Chang17}, we show that the
  4B5B encoding scheme can be used for constructing a strong ternary symmetrization mapping.
 In such an encoding scheme, each piece of 4 bits is uniquely mapped to a 5-bit codeword (see Table \ref{table:4B5B}).
 One salient feature of the 4B5B encoding scheme is that each 5-bit codeword has at most one leading 0 as well as at most two trailing 0's.
Thus, encoding the $L$-bit integer results in a $\lceil L/4 \rceil *5$-bit codeword that does not have 4 consecutive 0's.
Instead of adding the 6-bit delimiter 100001 in \cite{Chang17}, we add the 6-trit delimiter 200001 in front of the $\lceil L/4 \rceil *5$-trit codeword to construct an $M=\lceil L/4 \rceil *5+6$ codeword.
The details of the mapping from an $L$-bit integer to an $M$-trit codeword is shown in Algorithm \ref{alg:4B5B}.

\begin{algorithm}\caption{The 4B5B strong ternary symmetrization mapping}\label{alg:4B5B}
%\begin{algorithmic}[1]
\KwIn{An integer $0 \le x \le 2^L-1$.}
\KwOut{An $M$-trit codeword $\big(w(0), w(1), \ldots, w(M-1)\big)$, where $M=\lceil L/4 \rceil *5+6$.}

\noindent 1: Let $\big (\betab_1(x),\betab_2(x), \ldots, \betab_{L}(x)\big )$ be the binary representation of $x$, i.e., $x=\sum_{i=1}^L \betab_i(x)2^{i-1}$.
  If $L$ is not an integer multiple of 4, append $4-(L\;\mbox{mod}\;4)$ 0's to the binary representation of $x$ to form a $\lceil L/4 \rceil *4$-bit binary vector.

\noindent 2: Use the 4B5B encoding scheme to encode the $\lceil L/4 \rceil *4$-binary vector  into a  $\lceil L/4 \rceil *5$-bit codeword.

\noindent 3:  Add the 6-trit delimiter 200001 in front of the $\lceil L/4 \rceil *5$-bit codeword to form a $(\lceil L/4 \rceil *5+6)$-trit codeword.
%\end{algorithmic}
\end{algorithm}

In the following lemma, we show that the 4B5B mapping in Algorithm \ref{alg:4B5B} is indeed a strong ternary symmetrization mapping.
Though the change from the 6-bit delimiter 100001 in \cite{Chang17} to the 6-trit delimiter 200001 in this paper seems to be small,
we note that the proof of \rlem{4B5B} is quite different from that in \cite{Chang17}.

\blem{4B5B}
For $L \ge 1$, the 4B5B mapping in Algorithm \ref{alg:4B5B} is indeed a strong ternary $M$-symmetrization mapping with $M=\lceil L/4 \rceil *5+6$.
\elem

\bproof
Since the first element in the 6-trit delimiter 200001 is 2, we know that $w_i(0)=2$ for all $i$.
From the 4B5B encoding scheme, we know that
the substring of $4$ consecutive 0's only appears in the $6$-trit delimiter  and thus it appears exactly once in the $M$-trit cyclically shifted codeword $\big(w(d), w(d+1), \ldots, w((M-1+d)\;\mbox{mod}\;M)\big)$ for any integer $0 \le d \le M-1$.
Now consider the codeword $\big(w_i(0), w_i(1), \ldots, w_i(M-1)\big)$ and the cyclically shifted codeword $\big(w_j(d), w_j(d+1), \ldots, w_j((M-1+d)\;\mbox{mod}\;M)\big)$.

\noindent {\em Case 1}. $(d\;\mbox{mod}\;M) = 0$ and $i \ne j$:
In this case,  the $6$-trit delimiters of two $M$-trit codewords are aligned.
Choose $\tau_1=5$ and we have  $w_i(\tau_1)=w_j(\tau_1)=w_j((\tau_1+d)\;\mbox{mod}\;M)=1$.
Since $i \ne j$,  we have from the one-to-one mapping of the 4B5B encoding scheme
that there exists $6 \le \tau_2 \le M-1$ such that
$w_i(\tau_2)\ne w_j(\tau_2) =w_j((\tau_2+d)\;\mbox{mod}\;M)$.
Thus, the condition (ii) in \rdef{strong} is satisfied.

\noindent {\em Case 2}. $(d\;\mbox{mod}\;M) =1,2,3,4$:

Let $k=(d\;\mbox{mod}\;M)$. Choose $\tau_1=5$ and we have $w_i(\tau_1)=w_i(5)=1$.
Also, choose $\tau_2=5-k$ and we have $w_i(\tau_2)=0$ and $w_j((\tau_2+d)\;\mbox{mod}\;M)=w_j(5)=1$.
Since we assume that $L \ge 1$, we know that $M \ge 11$. Thus, $(\tau_1+d)\;\mbox{mod}\;M \ne 0$ and  $w_j((\tau_1+d)\;\mbox{mod}\;M) \ne 2$.
This then implies that $w_j((\tau_1+d)\;\mbox{mod}\;M)$ is either 0 or 1.
If $w_j((\tau_1+d)\;\mbox{mod}\;M)=0$, then condition (i) in \rdef{strong} is satisfied. On the other hand, if $w_j((\tau_1+d)\;\mbox{mod}\;M)=1$, the condition (ii) in \rdef{strong} is satisfied.

\noindent {\em Case 3}. $(d\;\mbox{mod}\;M) =M-1,M-2,M-3,M-4$:

This is the same as Case 2 once we interchange $i$ and $j$.

\noindent {\em Case 4.} $(d\;\mbox{mod}\;M) =5$:

In this case, Choose $\tau_1=5$ and we have $w_i(\tau_1)=1$. Since we assume that $L \ge 1$, we know that $M \ge 11$ and thus $w_j(10) \ne 2$.
This then implies that $w_j(10)=w_j((\tau_1+d)\;\mbox{mod}\;M)$ is either 0 or 1.
 Note that in this case we also have $w_i(1)=\ldots=w_i(4)=0$ and $w_j((t+5)\;\mbox{mod}\;M)$, $t=1,\ldots,4$, cannot be all 0's. Thus, there exists $1 \le \tau_2 \le 4$ such that $w_i(\tau_2)=0$ and $w_j((\tau_2+d)\;\mbox{mod}\;M)=1$. If  $w_j(10)=0$, the condition (i) in \rdef{strong} is satisfied. On the other hand, if $w_j(10)=1$, the condition (ii) in \rdef{strong} is satisfied.

\noindent {\em Case 5.} $(d\;\mbox{mod}\;M) =M-5$:

This is the same as Case 4 once we interchange $i$ and $j$.

\noindent {\em Case 6}. $(d\;\mbox{mod}\;M)$ is not in $\{M-5,M-4,M-3,M-2,M-1,0,1,2,3,4,5\}$:

In this case, the $6$-trit delimiters of the two $M$-trit codewords do not overlap. Then we have $w_i(1)=\ldots=w_i(4)=0$ and $w_j((t+d)\;\mbox{mod}\;M)$, $t=1,\ldots,4$, cannot be all 0's. Thus, there exists $1 \le \tau_2 \le 4$ such that $w_i(\tau_2)=0$ and $w_j((\tau_2+d)\;\mbox{mod}\;M)=1$. On the other hand, we have $w_j(1)=\ldots=w_j(4)=0$ and $w_i((t-d)\;\mbox{mod}\;M)$, $t=1,\ldots,4$, cannot be all 0's. Thus, there exists $1 \le ((\tau_1+d)\;\mbox{mod}\;M) \le 4$ such that $w_i(\tau_1)=1$ and $w_j((\tau_1+d)\;\mbox{mod}\;M)=0$.

\eproof

\bsubsec{The quasi-random algorithm}{modular}

In this section, we propose the quasi-random algorithm for guaranteed rendezvous in the symmetric, anonymous, asynchronous, and heterogeneous setting, where each SU has a single radio. Our quasi-random algorithm is an extension of the construction in \cite{Chang17} without the need of the unique ID assumption.
For this, we first introduce the  modular clock algorithm in \cite{Theis2011} (see Algorithm \ref{alg:clock}).
 In addition to the available channel set, the algorithm needs three parameters: the period $p$ that is an integer not smaller than the number of available channels $n$, the slope $r$ that is relatively prime to $p$, and the bias that is an integer selected from $\{0,1, \ldots, p-1\}$. If the clock $k$ in Line 3 of the algorithm is not greater than $n-1$, then a channel is selected at random from the available channel set.

\iffalse
\begin{algorithm}\caption{The deterministic modular clock algorithm}\label{alg:clock}
\begin{algorithmic}[1]
\Require  An available channel set ${\bf c}=\{c(0), c(1), \ldots, c(n-1)\}$, a period $p \ge n$, a slope $r>0$ that is relatively prime to $p$, and a bias $0 \le b \le p-1$.
\Ensure A deterministic sequence $\{\alpha(t), t=0,1,\ldots \}$ with $\alpha(t) \in {\bf c}$.
\State Let $z=0$.
\State For each $t$, let $k=((r*t+b)\;\mbox{mod}\;p)$.
\State If $k\le n-1$, let $\alpha(t)=c(k)$.
\State Otherwise, let $\alpha(t)=c(z)$ and update $z \leftarrow ((z+1)\;\mbox{mod}\;n)$.
 \end{algorithmic}
\end{algorithm}
\fi

\begin{algorithm}\caption{The  modular clock algorithm}\label{alg:clock}
%\begin{algorithmic}[1]
%\noindent {\bf Input}
\KwIn{An available channel set ${\bf c}=\{c(0), c(1), \ldots, c(n-1)\}$, a period $p \ge n$, a slope $r>0$ that is relatively prime to $p$,  a bias $0 \le b \le p-1$, and an index of time $t$.}
%, a slope $r>0$ that is relatively prime to $p$, and a bias $0 \le b \le p-1$.

%\noindent {\bf Output}
\KwOut{A channel $\alphax(t) \in {\bf c}$.}

%\noindent  Let $z=0$.

\noindent 1: For each $t$, let $k=((r*t+b)\;\mbox{mod}\;p)$.

\noindent 2: If $k\le n-1$, let $\alphax(t)=c(k)$.

\noindent 3: Otherwise, select $\alphax(t)$ uniformly at random  from the available channel set ${\bf c}$.

%\end{algorithmic}
\end{algorithm}

One well-known property of the modular clock algorithm in Algorithm \ref{alg:clock} is the rendezvous property from the Chinese Remainder Theorem.

\bprop{clock}(Theorem 4 of \cite{Theis2011})
Suppose that user $1$ (resp. user $2$) uses the  modular clock algorithm in Algorithm \ref{alg:clock} to generate its CH sequence
with the period $p_1$ (resp. $p_2$). If $p_1$ and $p_2$ are relatively prime, then under the assumption in \req{avail1111}, these two users will rendezvous  within $p_1 p_2$ time slots.
\eprop

Now we combine the  modular clock algorithm in Algorithm \ref{alg:clock} and the strong ternary symmetrization mapping in Algorithm \ref{alg:4B5B} to construct a CH that can provide guaranteed rendezvous. Such an algorithm is called the {\em quasi-random algorithm} in this paper and its detail is shown
 in Algorithm \ref{alg:hop}. The idea is to randomly select a channel $c$ (as its ID) from the available channel set and then map $c$ to an $M$-trit codeword by the 4B5B  strong ternary symmetrization mapping
 in Algorithm \ref{alg:4B5B}.
 Then we interleave $M$ sequences according to the ternary value of its $M$-trit  codeword. Specifically, for user $i$, we select two {\em primes} $p_{i,0}$ and $p_{i,1}$ such that $n_i \le p_{i,0} < p_{i,1}$.
 A 0-sequence (resp. 1-sequence) of user $i$ is then constructed by using the  modular clock algorithm with the prime $p_{i,0}$ (resp. $p_{i,1}$).
 The slope parameter and the bias parameter are selected at random.
 A 2-sequence is a ``stay'' sequence in which channel $c$ is used in every time slot. Then the CH sequence of a user is constructed by interleaving $M$ $\{0/1/2\}$-sequences according to its $M$-trit codeword.
 Let $\{X_i(t), t\ge 0\}$ be the CH sequence for user $i$, $i=1$ and 2.
 The insight behind our construction is that the two users will rendezvous immediately at time 0 during the 2-sequence if  both users select the same channel as their IDs
and their clocks are synchronized. On the other hand, either their clocks are not synchronized or their IDs are different, the strong ternary symmetrization mapping  in \rdef{strong} ensures that there exists some time $\tau$ such that the subsequence $\{\alphax_1(\tau),\alphax_1(\tau+M), \alphax_1(\tau+2M), \ldots \}$
and the subsequence $\{\alphax_2(\tau),\alphax_2(\tau+M), \alphax_2(\tau+2M), \ldots \}$
are generated by the modular clock algorithm with two different primes. These two users are then guaranteed to rendezvous from the Chinese Remainder Theorem for the modular clock algorithm in \rprop{clock}. The result and the detailed proof is shown in the following theorem.

\begin{algorithm}\caption{The quasi-random algorithm}\label{alg:hop}
%\begin{algorithmic}[1]
\KwIn{An available channel set ${\bf c}=\{c(0), c(1), \ldots, c(n-1)\}$ and the total number of channels $N$.}
\KwOut{A CH sequence $\{\betax(t), t=0,1,\ldots \}$ with $\betax(t) \in {\bf c}$.}

\noindent 1: Randomly select a channel $c$ from the available channel set.
Use the 4B5B $M$-symmetrization mapping (Algorithm \ref{alg:4B5B}) to map $c$ to an $M$-trit codeword $(w(0),w(1), \ldots, w(M-1))$ with $M=\lceil \lceil \log_2 N \rceil /4 \rceil *5+6$.

\noindent 2: Select two primes $p_1 >p_0 \ge n$.

\noindent 3: For each $s=1,2,\ldots, M-1$, generate independent  and uniformly distributed random variables $r_0(s) \in [1,p_0-1]$,
$r_1(s) \in [1,p_1-1]$, $b_0(s) \in [0,p_0-1]$ and $b_1(s) \in [0,p_1-1]$.

\noindent 4: For each $t$, compute the following two variables:

\noindent 5: $q=\lfloor t/M\rfloor$.

\noindent 6: $s =(t \;\mbox{mod}\; M)$.

\noindent 7: If $w(s)=2$, let  $\betax(t)=c$.

\noindent 8: If $w(s)=1$, let $\betax(t)$ be the output channel from the  modular clock algorithm in Algorithm \ref{alg:clock}
with the period $p_1$, the slope $r_1(s)$,  the bias $b_1(s)$, and the index of time $q$.

\noindent 9: If $w(s)=0$, let $\betax(t)$ be the output channel from the  modular clock algorithm in Algorithm \ref{alg:clock}
with the period $p_0$, the slope $r_0(s)$,  the bias $b_0(s)$, and the index of time $q$.
% \end{algorithmic}
\end{algorithm}

%Thus, its ETTR performance should be comparable to that of the random algorithm (if $M$ is not too small). In \rsec{exp}, we will show by computer %simulations that the ETTR of our algorithm for two-user rendezvous is almost the same as that of the random algorithm.

 %We note that it is possible that $p_{1,0}=p_{2,1}$ and thus the previous relatively prime argument for interleaving $M$ 0/1-sequences according to cyclic unique codewords fails. Fortunately, the two properties  for a strong symmetrization mapping is much stronger than the cyclic unique property and we can use them to prove guaranteed rendezvous in the following theorem.

 \bthe{main4B5B}(The MTTR bound)
 Suppose the assumption in  \req{avail1111} hold and  the two users use the quasi-random algorithm in  Algorithm \ref{alg:hop} to generate their CH sequences. Then
 these two users will rendezvous  within $M p_{1,1} p_{2,1}$ time slots, where $M=\lceil \lceil \log_2 N \rceil /4 \rceil *5+6$ and $N$ is the total number of channels.
 \ethe

Since there are two primes between $[n, 3n]$ \cite{Prime06},  these two users will rendezvous  within $9M n_1n_2$ time slots.

\bproof
Let $d$ be the clock shift between these two users. Suppose that user 1 (resp. 2) selects $c_1$ (resp. $c_2$) to construct its codeword.
Note from  Algorithm \ref{alg:hop} that
for $t \in \{\tau, \tau+M, \tau+2M, \ldots\}$, user $1$ uses a $w_1(\tau)$-sequence and user $2$ uses a $w_2(\tau+d)$-sequence.
Let $X_i(t)$, $i=1$ and 2, be the channel selected by user $i$ at time $t$.
In view of the definition of a strong ternary symmetrization mapping in \rdef{strong},  we consider the following two cases.

\noindent  {\em Case 1}. $(d\;\mbox{mod}\;M) = 0$ and $c_1=c_2$:

From Step 1 of Algorithm \ref{alg:hop}, these two users use the same codeword. Since $(d\;\mbox{mod}\;M) = 0$, the 6-trit delimiters of these two users are aligned. Thus, for $t \in \{0, M, 2M, \ldots\}$, we know from Step 7 of Algorithm \ref{alg:hop},   user 1 (resp. 2) stays on channel $c_1$
(resp. $c_2$). Since $c_1=c_2$, both users rendezvous at time 0.

\noindent  {\em Case 2}. $(d\;\mbox{mod}\;M) \ne 0$:

There are two subcases.

\noindent {\em Case 2.1}. There exist $0 \le \tau_1, \tau_2 \le M-1$ such that $w_1(\tau_1)=1, w_2((\tau_1+d)\;\mbox{mod}\;M)=0$ and
$w_1(\tau_2)=0, w_2((\tau_2+d)\;\mbox{mod}\;M)=1$:

In this case, for $t \in \{\tau_1, \tau_1+M, \tau_1+2M, \ldots\}$, user $1$ uses a 1-sequence and user $2$ uses a 0-sequence.
The 1-sequence of user $1$ is generated from the  modular clock algorithm with the prime $p_{1,1}$
and the 0-sequence of user $2$ is generated from the  modular clock algorithm with the prime $p_{2,0}$.
If $p_{1,1} \ne p_{2,0}$, then we conclude from \rprop{clock} that these two users will rendezvous  within $M p_{1,1} p_{2,0}$ time slots.

On the other hand, if $p_{1,1}=p_{2,0}$, then we have
$$p_{2,1}> p_{2,0}=p_{1,1}> p_{1,0}.$$
Now for $t \in \{\tau_2, \tau_2+M, \tau_2+2M, \ldots\}$, user $1$ uses a 0-sequence and user $2$ uses a 1-sequence.
The 0-sequence of user $1$ is generated from the  modular clock algorithm with the prime $p_{1,0}$
and the 1-sequence of user $2$ is generated from the  modular clock algorithm with the prime $p_{2,1}$.
Since $p_{2,1} \ne p_{1,0}$, we know from \rprop{clock} that these two users will rendezvous  within $M p_{1,0} p_{2,1}$ time slots.

\noindent {\em Case 2.2}. There exist $0 \le \tau_1, \tau_2 \le M-1$ such that $w_1(\tau_1)=w_2((\tau_1+d)\;\mbox{mod}\;M)=1$,
and $w_1(\tau_2)\ne w_2((\tau_2+d)\;\mbox{mod}\;M)$:

In this case,
for $t \in \{\tau_1, \tau_1+M, \tau_1+2M, \ldots\}$, user $1$ uses a 1-sequence and user $2$ uses a 1-sequence.
The 1-sequence of user $1$ is generated from the modular clock algorithm with the prime $p_{1,1}$ and
the 1-sequence of user $2$ is generated from the  modular clock algorithm with the prime $p_{2,1}$.
If  $p_{1,1} \ne p_{2,1}$, then we conclude from \rprop{clock} that these two users will rendezvous  within $M p_{1,1} p_{2,1}$ time slots.

On the other hand, if $p_{1,1}=p_{2,1}$, then we have
\bear{4B5B1111}
&&p_{2,1}=p_{1,1}> p_{1,0}, \nonumber \\
&&p_{1,1}=p_{2,1}> p_{2,0}.
\eear
Now for $t \in \{\tau_2, \tau_2+M, \tau_2+2M, \ldots\}$,  user $1$ uses a $w_1(\tau_2)$-sequence and user $2$ uses a $w_2(\tau_2+d)$-sequence with $w_1(\tau_2) \ne w_2(\tau_2+d)$. In view of \req{4B5B1111}, we conclude from \rprop{clock} that these two users will also rendezvous  within $M \max[p_{1,0} p_{2,1}, p_{1,1} p_{2,0}]$ time slots in this case.
\eproof

\iffalse
Since there is a prime between $[n, 2n]$ \cite{Erdos32} and another prime in $[2n,3n]$ \cite{Prime06}, we then have the following corollary.

\bcor{main4B5B}
Suppose the assumption in  \req{avail1111} hold and  the two users use Algorithm \ref{alg:hop} to generate their CH sequences.
Then these two users will rendezvous within $6 M n_1 n_2$ time slots, where $M=\lceil \lceil \log_2 N \rceil /4 \rceil *5+6$ and $N$ is the total number of channels.
\ecor
\fi

As commented in \cite{Chang17},   one way to reduce the ETTR is to avoid introducing ``stay'' modes that repeatedly examine the same channel pairs of two users. As such, the slope $r$ chosen in Line 3 of the algorithm  is an integer in $[1,p-1]$ and it is selected independently for $s=1,2,\ldots, M-1$. As the slope $r$ is nonzero, there is no ``stay'' mode in this algorithm except the case $s=0$ (with $w(0)=2$). On the other hand, the bias $b$ chosen in Line 3 of the algorithm is an integer in $[0, p-1]$.  For $s=0$, we have $w(0)=2$ and the quasi-random algorithm outputs the randomly selected channel $c$ from the available channel set. Since all the slopes and biases for $s=1,2, \ldots, M-1$ are generated independently and uniformly,
it is straightforward to verify that  the quasi-random algorithm selects each available channel independently with an equal probability in
the first $M$ time slots, i.e.,
$\{\alphax(t), t=d,d+1,d+2,\ldots, d+M-1\}$ are independently and identically distributed (i.i.d.) random variables with $\pr(\alphax(t)=c(\ell))={1}/{n}$ for all $\ell=0,1,\ldots, n-1$.
As such, the quasi-random algorithm behaves as if it were a random algorithm for every consecutive $M$ slots.
On the other hand, $\alphax(t)$ and $\alphax(t+qM)$ are correlated through the modular clock algorithm  as they both have the same value of $s$ and thus the same  slope $r$ and bias $b$. Such a correlated property ensures that the MTTR is bounded as shown in \rthe{main4B5B}.
In the following theorem, we use the i.i.d. property and the MTTR bound in \rthe{main4B5B} to derive an ETTR bound for the quasi-random algorithm.

%Analogous to  the argument for the ETTR bound in (10) of \cite{Chang17},
% we can also use the i.i.d. property and the MTTR bound in \rthe{main4B5B} to derive an ETTR bound for the quasi-random algorithm.
% This is stated in the following theorem.

 \bthe{main4B5Bb}(The ETTR bound)
 Suppose the assumption in  \req{avail1111} hold and  the two users use the quasi-random algorithm in Algorithm \ref{alg:hop} to generate their CH sequences.
 Then the ETTR is upper bounded by
\beq{ETTR1234}
\frac{n_1 n_2}{G}+9Mn_1n_2 \cdot (1-\frac{G}{n_1 n_2})^M,
\eeq
where $M=\lceil \lceil \log_2 N \rceil /4 \rceil *5+6$, $N$ is the total number of channels, and $G$ is the number of common channels between these two users.
  \ethe

Note that the first term in \req{ETTR1234} is the ETTR of the random algorithm.
Clearly, the second term in \req{ETTR1234} converges to 0 as $M \to \infty$. Thus, the ETTR of the quasi-random algorithm is almost the same as that of the random algorithm when $M$ is large.
On the other hand, if $M$ is very small, then the ETTR bound in \rthe{main4B5Bb} could be much larger than the ETTR  of the random algorithm. Also, as $M$ is very small, the quasi-random algorithm will hop to the ID channel very often and this might, in fact, increase the ETTR if the ID channel is not a rendezvous channel. As such, for the practical use of the quasi-random algorithm, one should avoid using a very small $M$. One easy way to do this to repeat the $L$-bit binary representation for several times in Step 1 of Algorithm \ref{alg:4B5B}. Or better yet, one may add a random binary vector in front of the $L$-bit binary representation to protect the user from jamming attack.
However, we note that increasing $M$ also increases the (theoretical) MTTR bound in \rthe{main4B5B}.

%\iffalse
\bproof
%Now we use these two properties to bound the ETTR of the quasi-random algorithm.
The proof of this theorem is similar to the argument for the ETTR bound in (10) of \cite{Chang17}.
Let $h=G/n_1n_2$ be the  probability that the two users hop on one common available channel by using the random algorithm. Clearly, the ETTR of the random algorithm is $1/h$. Also, let $H=9Mn_1n_2$ be the upper bound for MTTR in \rthe{main4B5B}.
Since  each user selects a channel independently and uniformly from its available channel set in the first $M$ time slots of the quasi-random algorithm, we then have
\bearn
\ex[T]&=&\sum_{t=1}^H t\cdot \pr(T=t)\nonumber \\
%&=&\sum_{t=1}^M t\cdot  \pr(T=t)+\sum_{t=M+1}^H t\cdot  \pr(T=t)\nonumber \\
&=&\sum_{t=1}^M t\cdot h(1-h)^{t-1}+\sum_{t=M+1}^H t\cdot  \pr(T=t)\nonumber\\
%&\leq&\sum_{t=1}^\infty t\cdot h(1-h)^{t-1}+H \sum_{t=M+1}^H  \pr(T=t)\nonumber\\
&\leq&\sum_{t=1}^\infty t\cdot h(1-h)^{t-1}+H \cdot  \pr(T>M)\nonumber\\
%&=& 1/h+H \cdot  \pr(T>M)\nonumber\\
&=& 1/h+H\cdot (1-h)^M.
\eearn

\eproof
%\fi

\begin{figure*}[!t]
\centering
\includegraphics[width=150mm]{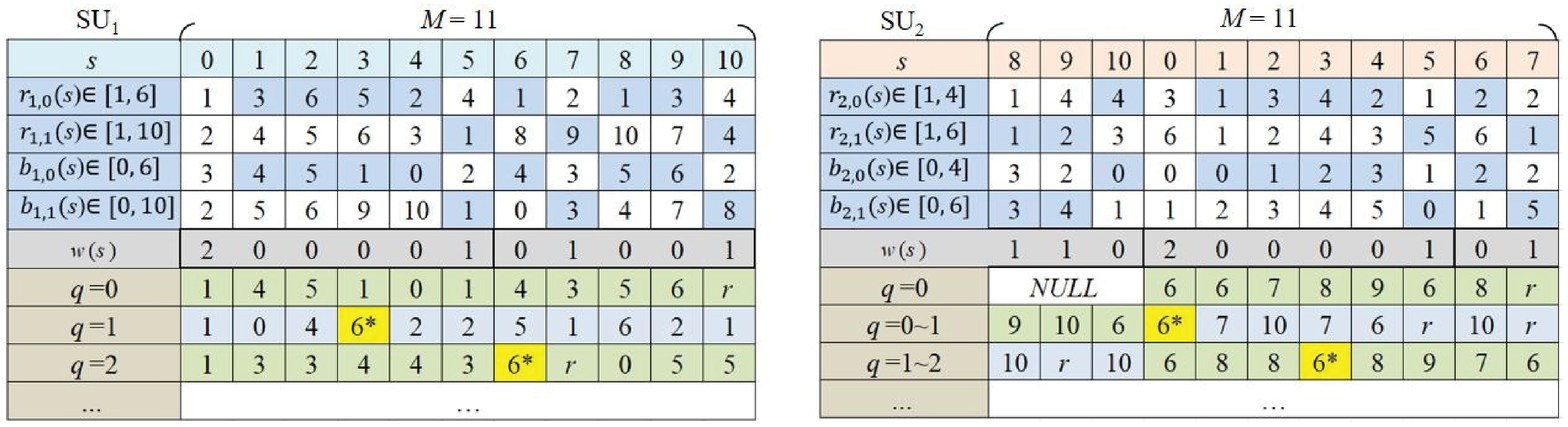}
 \\\footnotesize{*: the channel where the two users rendezvous.}
  \\\footnotesize{$r$: a channel replaced by a randomly chosen channel from the available channel set.}
    \caption{An illustrating example of the quasi-random algorithm with two users.
    }
    \label{fig:my3}
\end{figure*}

In \rfig{my3}, we provide an illustrating example for the constructions of  the CH sequences of Algorithm \ref{alg:hop} for a CRN with two users,
$SU_1$ and $SU_2$.
%To illustrate the effect that the clocks of these two users are not synchronized,
In this example, we assume that there is a clock drift of three time slots between these two users.
Suppose that there are $N$ = 15 channels, and each user has a single radio, i.e., $m_1$ = $m_2$ = 1. The available channels for $SU_1$ is $\{0, 1, 2, 3, 4, 5, 6\}$ and the available channels for $SU_2$ is $\{6, 7, 8, 9, 10\}$.
Thus, $n_1$ = 7 and $n_2$ = 5. Thus, we can simply choose $p_{1,0} = 7$, $p_{1,1} = 11$, $p_{2,0} = 5$, and $p_{2,1} = 7$.
Suppose that
$SU_1$ randomly selects a channel $c=1$  from the available channel set (as its ID).
From Table \ref{table:4B5B}, we know the 5B code for 1 (i.e. 0001) is 01001.
According to the  4B5B strong ternary symmetrization mapping in Algorithm \ref{alg:4B5B}, we then add the 6-trit delimiter 200001
in front of the 5B code 01001.
This then leads to the $11$-trit codeword
$$(w(0), w(1),\ldots, w(10))=(2, 0, 0, 0, 0, 1, 0, 1, 0, 0, 1).$$ For each $s = 0, 1,\ldots, 10$, we generate independent and uniformly distributed random variables $r_{1, 0}(s)\in{\bf[1, 6]}, r_{1, 1}(s)\in{\bf[1, 10]}, b_{1, 0}(s)\in{\bf[0, 6]}, b_{1, 1}(s)\in{\bf[0, 10]}$ and these are shown in \rfig{my3}.
Therefore, at $t=0$, we have $w(0)=2$. Thus, $\alphax(0)=c= 1$ (as shown in \rfig{my3}). Similarly, $\{\alphax(t), t=11, 22,\ldots\}$ are also  $c$ (i.e. 1).
 Now for $t \ne 0, 11,22, \ldots$, we compute $s=(t\; \mbox{mod}\;11)$ and $q=\lceil t/M \rceil$. If $w(s)=1$, we use $r_{1, 1}$ and $b_{1, 1}$ to generate the $\alphax(t)=c((r_{1, 1}*q+ b_{1, 1}) \;\mbox{mod}\; p_{1,1})$  if ($r_{1, 1}*q+ b_{1, 1}) \;\mbox{mod}\; p_{1,1}$ is not larger than $n_1$, i.e. 7. Otherwise, we randomly choose a channel from the available channel set (see e.g., $t=10$ in \rfig{my3}).
Similarly, if $w(s)=0$, we use $r_{1, 0}$ and $b_{1, 0}$ and $p_{1, 0}$ as the input of the modular clock algorithm to generate $\alphax(t)$.
For $SU_2$, suppose that it selects channel 6 (as its ID).
From the  4B5B strong ternary symmetrization mapping in Algorithm \ref{alg:4B5B}, its $11$-trit codeword is
$$(w(0), w(1),\ldots, w(10))=(2, 0, 0, 0, 0, 1, 0, 1, 1, 1, 0).$$
The CH sequences for $SU_2$ are generated similarly as shown in \rfig{my3}.
Note that  these two SUs will rendezvous on the common channel (i.e. channel 6) at  $t=14$ and $t=28$, respectively.

\bsubsec{Multi-radio CH sequences}{multiple}

Now we consider the multiple radio setting.
Suppose that user $i$ has $m_i \ge 1$ radios, $i=1$ and 2. It is possible that $m_i= 1$ in this setting.
As shown in \cite{RPS},  if we generate {\em independently} the channel hopping sequence for each radio by using  the single-radio algorithm in Algorithm \ref{alg:hop}, then
it fails to improve the MTTR bound by using multiple radios. To further improve the MTTR bound in the multiple radio setting, we follow the approach in \cite{ICC}. We first divide the $n_i$ available channels as evenly as possible to the $m_i$ radios so that each radio is assigned with at most $\lceil n_i/m_i\rceil$ channels. Let ${\bf c}_{i}^{(k)}$ be the channel assigned to the $k^{th}$ radio of user $i$. For the $k^{th}$ radio of user $i$, construct the CH sequence by using the single-radio algorithm in Algorithm \ref{alg:hop} with the input
${\bf c}_i^{(k)}$ and $N$. The detailed algorithm is shown in Algorithm \ref{alg:multiple}.

\begin{algorithm}[t]
\caption{The  multiple radio algorithm}\label{alg:multiple}
%\begin{algorithmic}[1]
%\noindent {\bf Input}
\KwIn{An available channel set ${\bf c}=\{c(0), c(1), \ldots, c(n-1)\}$, the number of radios $m$ and the total number of channels in the CRN $N$.}
% a period $p \ge n$, a slope $r>0$ that is relatively prime to $p$, and a bias $0 \le b \le p-1$.

%\noindent {\bf Output}
\KwOut{$m$ CH sequences with $\{\alphax^{(k)}(t), , t=0,1,\ldots \}$, $k=1,2, \ldots, m$ for the $k^{th}$ radio.}
 %with $\alphax(t) \in {\bf c}$.

%\noindent  Let $z=0$.
\noindent 1:  Assign the $|{\bf c}|$ channels in the round robin fashion  to the $m$ radios.
Let ${\bf c}^{(k)}$ be the set of channels assigned to the $k^{th}$ radio, $k=1,2,\ldots, m$.

\noindent 2: For the $k^{th}$ radio, construct the CH sequence by using the single-radio algorithm in Algorithm \ref{alg:hop} with the input
${\bf c}^{(k)}$ and $N$.

%\end{algorithmic}
\end{algorithm}

Analogous to the argument for Theorem 5 in \cite{ICC},
one can easily argue by contradiction that there must exist some $1 \le k_1^* \le m_1$ and $1 \le k_2^* \le m_2$ such that
$${\bf c}^{(k_1^*)}_1 \cap {\bf c}^{(k_2^*)}_2 \ne\phi.$$
Since $|{\bf c}^{(k_i^*)}_1| \le \lceil n_i/m_i\rceil$, $i=1$ and 2,  the MTTR bound for the multiple radio algorithm in
Algorithm \ref{alg:multiple} then follows directly from the MTTR bound for a single radio in \rthe{main4B5B}.
This is stated in the following corollary.

\bcor{MultipleRadio}
Suppose that
the assumption in \req{avail1111} holds. User $i$ uses the multiple radio algorithm in
Algorithm \ref{alg:multiple}  to
generate its CH sequence.
Then both users rendezvous within $9 M \lceil n_1/m_1 \rceil \cdot \lceil n_2/m_2 \rceil$ time slots, where
$$M=\lceil \lceil \log_2 N \rceil /4 \rceil *5+6.$$
% (with $N$ being the total number of channels).
\ecor

\iffalse
\bproof
%the proof of Theorem 5 in \cite{ICC}
Under the assumption in \req{avail1111},
one can easily argue by contradiction that there must exist some $1 \le k_1^* \le m_1$ and $1 \le k_2^* \le m_2$ such that
$${\bf c}^{(k_1^*)}_1 \cap {\bf c}^{(k_2^*)}_2 \ne\phi.$$
Since $|{\bf c}^{(k_i^*)}_1| \le \lceil n_i/m_i\rceil$, $i=1$ and 2,  the MTTR bound for the multiple radio algorithm in
Algorithm \ref{alg:multiple} then follows from the MTTR bound for a single radio in \rthe{main4B5B}.
\eproof
\fi

We note that the MTTR bound in \cite{ICC} is $O((\log \log N) \frac{n_1 n_2}{m_1 m_2})$. Our MTTR bound in \rcor{MultipleRadio} is only slightly larger than that
in \cite{ICC}.

We note the analysis for the ETTR for the quasi-random algorithm in the multi-radio setting is much more involved. For this, we will resort to computer simulations in \rsec{exp}.

 \bsec{Simulation Results}{exp}

In this section, we conduct extensive simulations to compare the performance of our quasi-random (QR) algorithm with several multi-radio channel hopping algorithms in asynchronous heterogenous CRNs, including the random algorithm, JS/I  \cite{RPS},  RPS \cite{RPS}, GCR \cite{Multiradio14},  AMRR/M (for optimizing MTTR) \cite{AMRR}, AMRR/E (for optimizing ETTR) \cite{AMRR}, and FMRR \cite{ICC}. Our simulations are performed with event-driven C++ simulators. The simulation setting is the same as that in \cite{ICC}.
Specifically, we assume that each SU is aware of its available channel set and the  total number of channels  $N$.
To model the clock drift, each user randomly selects a (local) time to start its CH sequence.
For each set of parameters, we generate 3,000 different available channel sets for the two users and perform 1,000 independent event-driven runs for each pair of the available channel sets.
We then compute the maximum/average time-to-rendezvous as the measured MTTR/ETTR. The simulation results are obtained with 95\% confidence intervals. Since the confidence intervals of ETTR are all very small in our simulations, for clarity, we do not draw the confidence intervals in the figures.

\bsubsec{Impact of the number of channels when the number of common channels is fixed}{fixed}

\iffalse
%one column
%\begin{UP}
\begin{figure}[htbp]
    \centering
    \begin{tabular}{p{0.45\textwidth}p{0.45\textwidth}}
    \includegraphics[width=0.45\textwidth]{sim1a.eps} &
    \includegraphics[width=0.45\textwidth]{sim1b.eps} \\
  (a) MTTR vs. the number of channels $N$  & (b) ETTR vs. the number of channels $N$ \\
    \end{tabular}
\caption{The effect of the number of channels on MTTR and ETTR  with  $n_1 = n_2$ uniformly chosen in $[14,16]$, $G = 2$ and $m_1 = 2, m_2 = 4$.}
    \label{fig:fig1}
\end{figure}
\fi

% Two columns
%\iffalse
\begin{figure}[htbp]
    \centering
    \includegraphics[width=80mm]{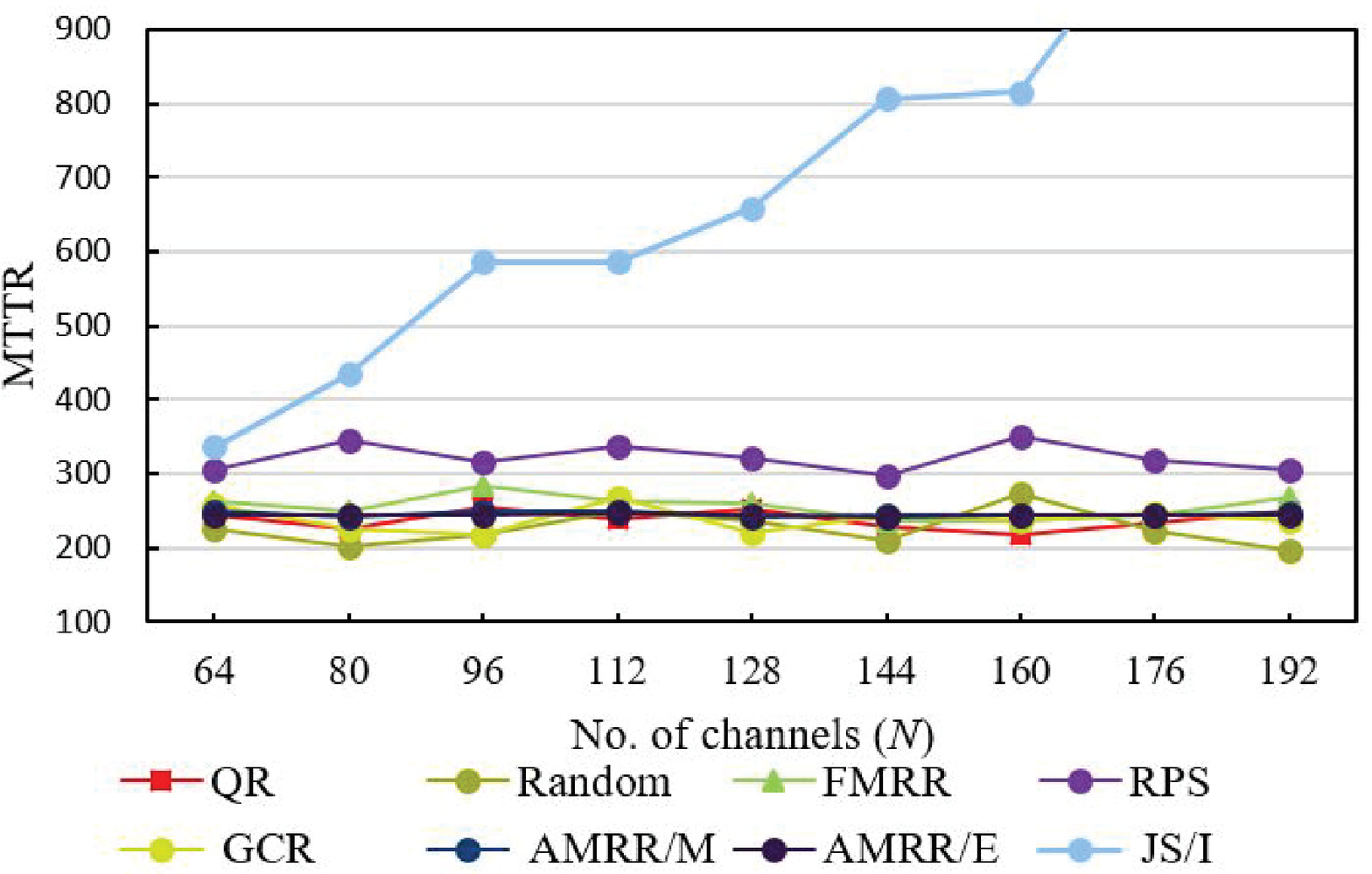}
 \\\footnotesize{ (a) MTTR vs. the number of channels $N$ }
    \includegraphics[width=80mm]{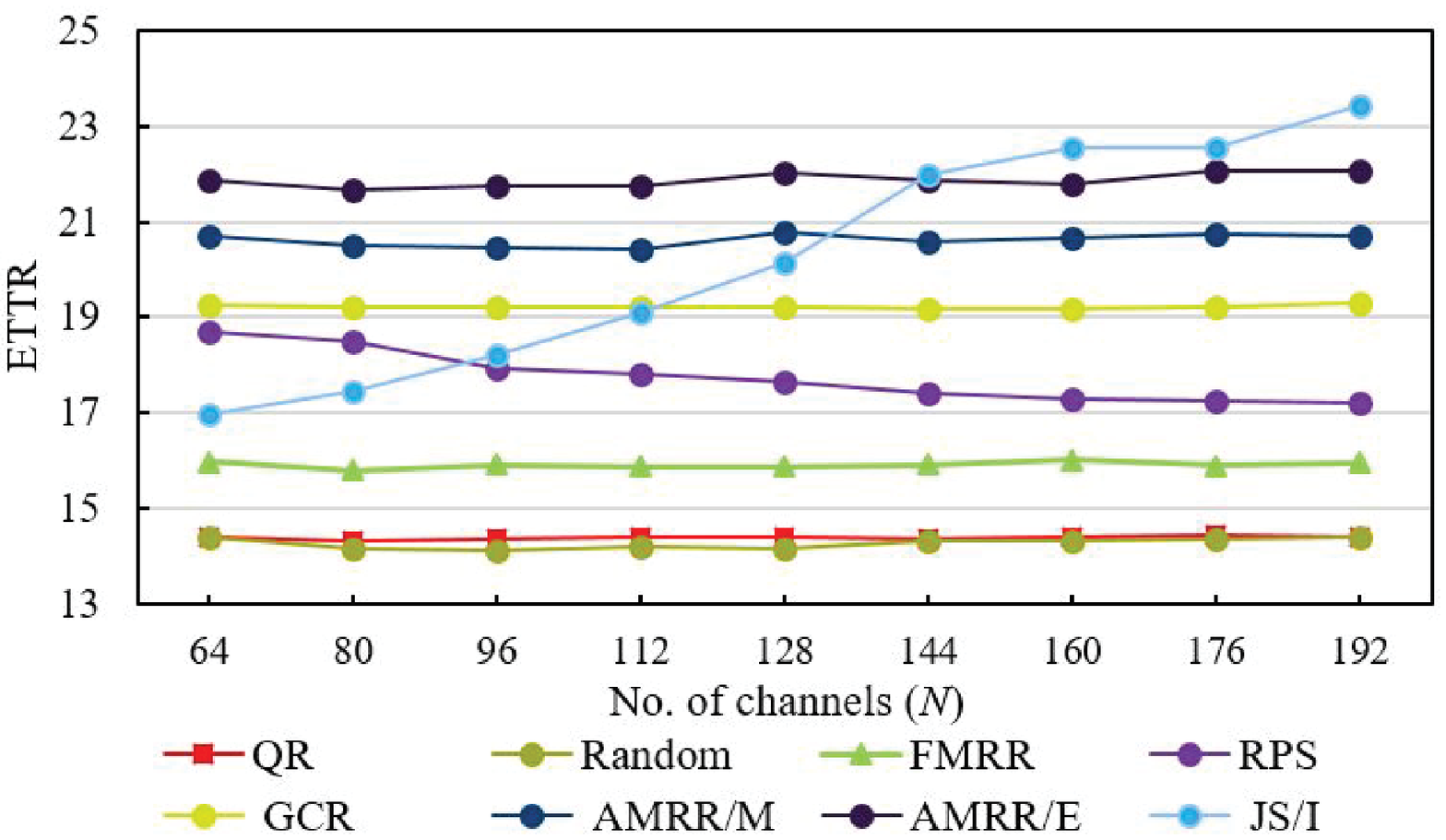}
 \\\footnotesize{ (b) ETTR vs. the number of channels $N$ }
\caption{The effect of the number of channels on MTTR and ETTR  with  $n_1 = n_2$ uniformly chosen in $[14,16]$, $G = 2$ and $m_1 = 2, m_2 = 4$.}
    \label{fig:fig1}
\end{figure}
%\fi

In this simulation, we vary the total number of channels $N$ from 64 to 192 with fixed $n_1 = n_2$ uniformly chosen in $[14,16]$,  $m_1 =2, m_2 = 4$, and the number of common channels $G = 2$.
In \rfig{fig1}(a), we show the MTTR results of all the algorithms. It is well-known
%(see e.g., Table I of \cite{ICC})
that the MTTR of
GCR \cite{Multiradio14} and that of AMRR/M \cite{AMRR} are $O(\frac{n_1 n_2}{m_1 m_2})$ (with the requirement that the number of radios for each user has to be larger than 1) and the MTTR of FMRR \cite{ICC} is $O(\frac{n_1 n_2}{m_1 m_2}\log(\log N))$. Even though the MTTR of our quasi-random (QR) algorithm is $O(\frac{n_1 n_2}{m_1 m_2}\log N)$ in theory,  the simulation results in \rfig{fig1}(a) show that
the MTTR of our algorithm is comparable to those of these three algorithms (i.e., GCR, AMRR/M, FMRR) and that of the random algorithm. Also, the MTTR of JS/I is $O(N^3)$ and its MTTR is significantly worse than the other algorithms.
%Moreover, the MTTR of our algorithm is almost invariant with respect to $N$ in the range from 80 to 128.

%An interesting outcome is when $n_1$ and $n_2$ are fixed, the MTTR of our algorithm is only affected by M (i.e. the number of time slots in an interval), O($\log\log N)$ . On the other hand, the MTTRs of JS/P and JS/I are $O(N^3)$. Additionally, when $N$ is large, the MTTRs of these algorithm is worse than ours.

In \rfig{fig1}(b), we show the ETTR results of all the algorithms. As shown in \rfig{fig1}(b), our algorithm performs much better than the other  schemes, and it is almost identical  to the ETTR of the random algorithm.
%
%Further, the ETTR of our algorithm remains same when the value of N varies from 144 to 192. The effect of the number of channels on MTTR and ETTR with $n_1 = n_2$ uniformly chosen in $[14,16]$, $G=2$,  $m_1 =2, m_2 = 4$.

%The main reason behind the excellent performance of our scheme is, even though JS, RPS and the proposed scheme hops randomly, the performance of %the former two schemes degrade as the value of $N$ increases. However, our ETTR remains same unaffected by the value of $N$.

\bsubsec{Impact of the number of channels when the number of common channels is proportional to the number of channels}{varying}

\iffalse
% one column
%\begin{UP}
\begin{figure}[htbp]
    \centering
    \begin{tabular}{p{0.45\textwidth}p{0.45\textwidth}}
    \includegraphics[width=0.45\textwidth]{sim2a.eps} &
    \includegraphics[width=0.45\textwidth]{sim2b.eps} \\
  (a) MTTR vs. the number of channels $N$  & (b) ETTR vs. the number of channels $N$ \\
    \end{tabular}
\caption{The effect of the number of channels on MTTR and ETTR  with  $n_1 = n_2 = N/2$, $G = N/8$ and $m_1 = 3, m_2 = 6$.}
    \label{fig:fig2}
\end{figure}
\fi

% Two columns
%\iffalse
\begin{figure}[htbp]
    \centering
    \includegraphics[width=80mm]{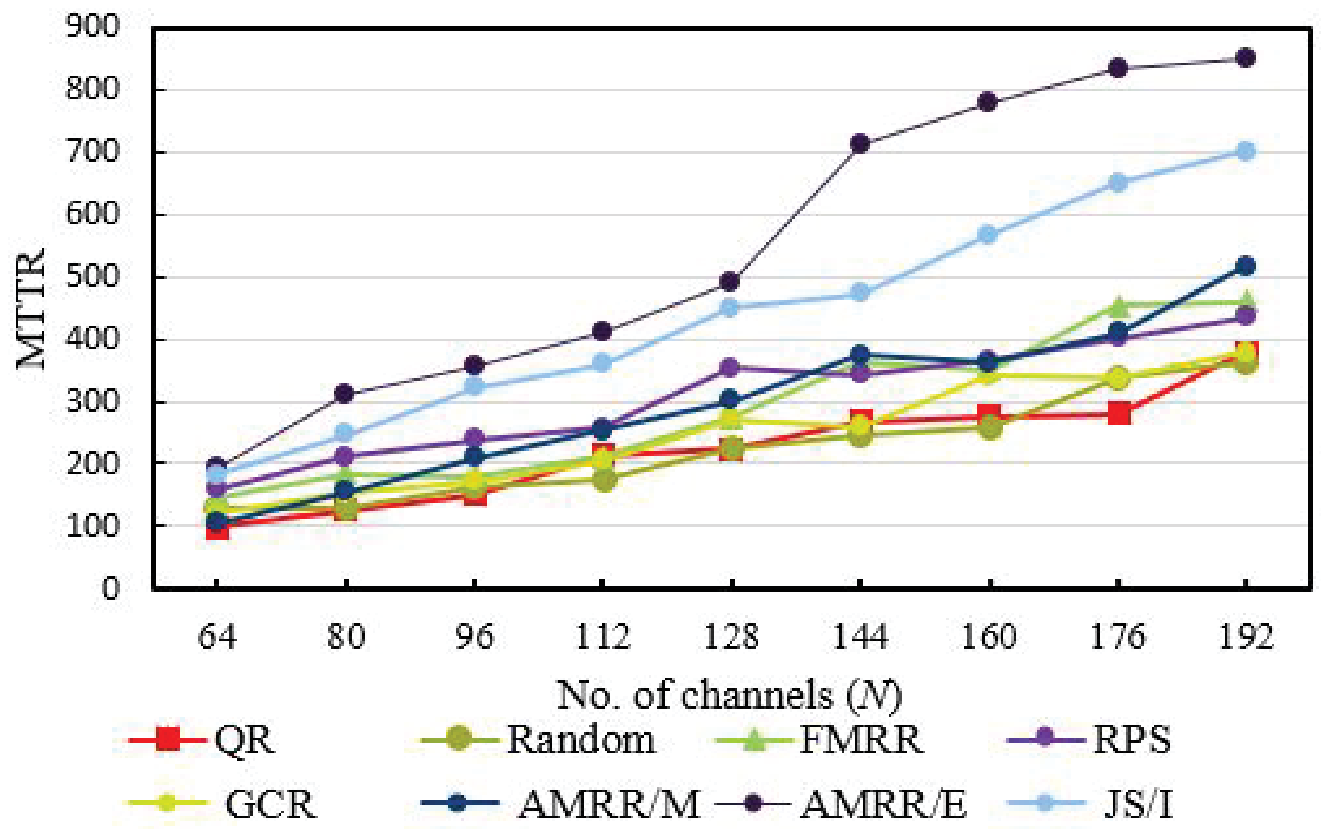}
 \\\footnotesize{ (a) MTTR vs. the number of total channels $N$ }
    \includegraphics[width=80mm]{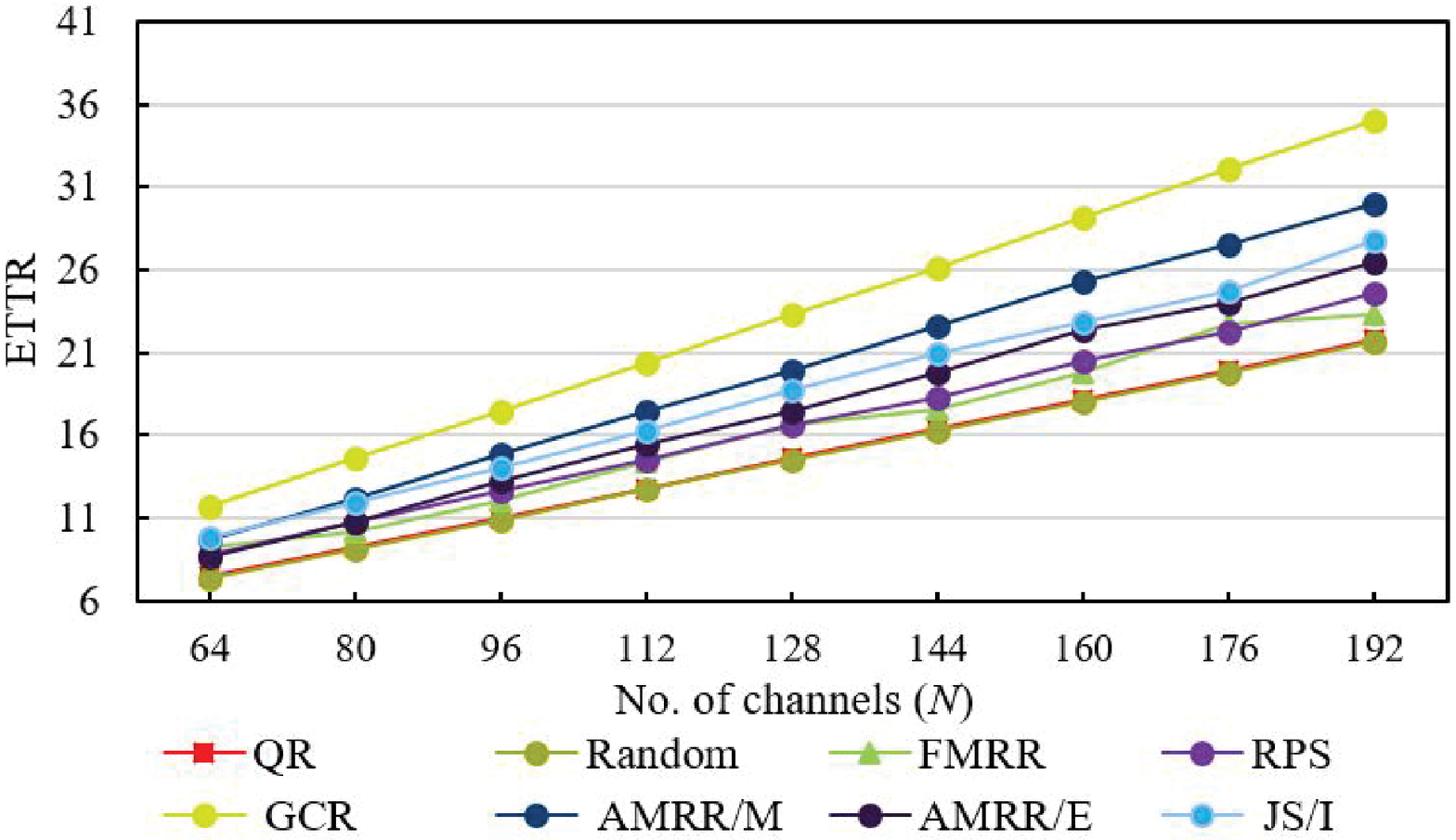}
 \\\footnotesize{ (b) ETTR vs. the number of total channels $N$ }
\caption{The effect of the number of channels on MTTR and ETTR  with  $n_1 = n_2 = N/2$, $G = N/8$ and $m_1 = 3, m_2 = 6$.}
    \label{fig:fig2}
\end{figure}
%\fi

In this simulation,  we vary $N$ from 64 to 192 and $n_1 = n_2 = N/2$, $G = N/8$ with fixed $m_1$ = 3 and $m_2 = 6$.
Since $n_1, n_2$ and $G$ are linear functions of $N$, it then follows from \rcor{MultipleRadio} that the MTTR of our QR algorithm is now $O(N^2 \log(N))$. As shown in  \rfig{fig2}(a), the MTTR of our algorithm is increasing in $N$, and it is also comparable to those of GCR, AMRR/M, FMRR and random algorithms.

%The MTTR of AMRR/M is smaller than that of AMRR/E as evident in the plot. Our MTTR is better than all the candidates.
In \rfig{fig2}(b), we show the ETTR results of all the algorithms in this simulation setting. Once again, our algorithm performs much better than the other  schemes, and it is almost identical  to the ETTR of the random algorithm.
%

%present the plot of ETTR for varying $N$. We can observe that our algorithm has the best ETTR due to its random hopping property.

\bsubsec{Impact of the number of radios}{Radios}

\iffalse
% one column
%\begin{UP}
\begin{figure}[htbp]
    \centering
    \begin{tabular}{p{0.45\textwidth}p{0.45\textwidth}}
    \includegraphics[width=0.45\textwidth]{sim3a.eps} &
    \includegraphics[width=0.45\textwidth]{sim3b.eps} \\
  (a) MTTR for various settings of $m_1$ and $m_2$  & (b) ETTR for various settings of  $m_1$ and $m_2$ \\
    \end{tabular}
\caption{The effect of the number of radios on MTTR and ETTR for various settings of $m_1$ and $m_2$.}
    \label{fig:fig3}
\end{figure}
\fi

% Two column
%\iffalse
\begin{figure}[htbp]
    \centering
    \includegraphics[width=80mm]{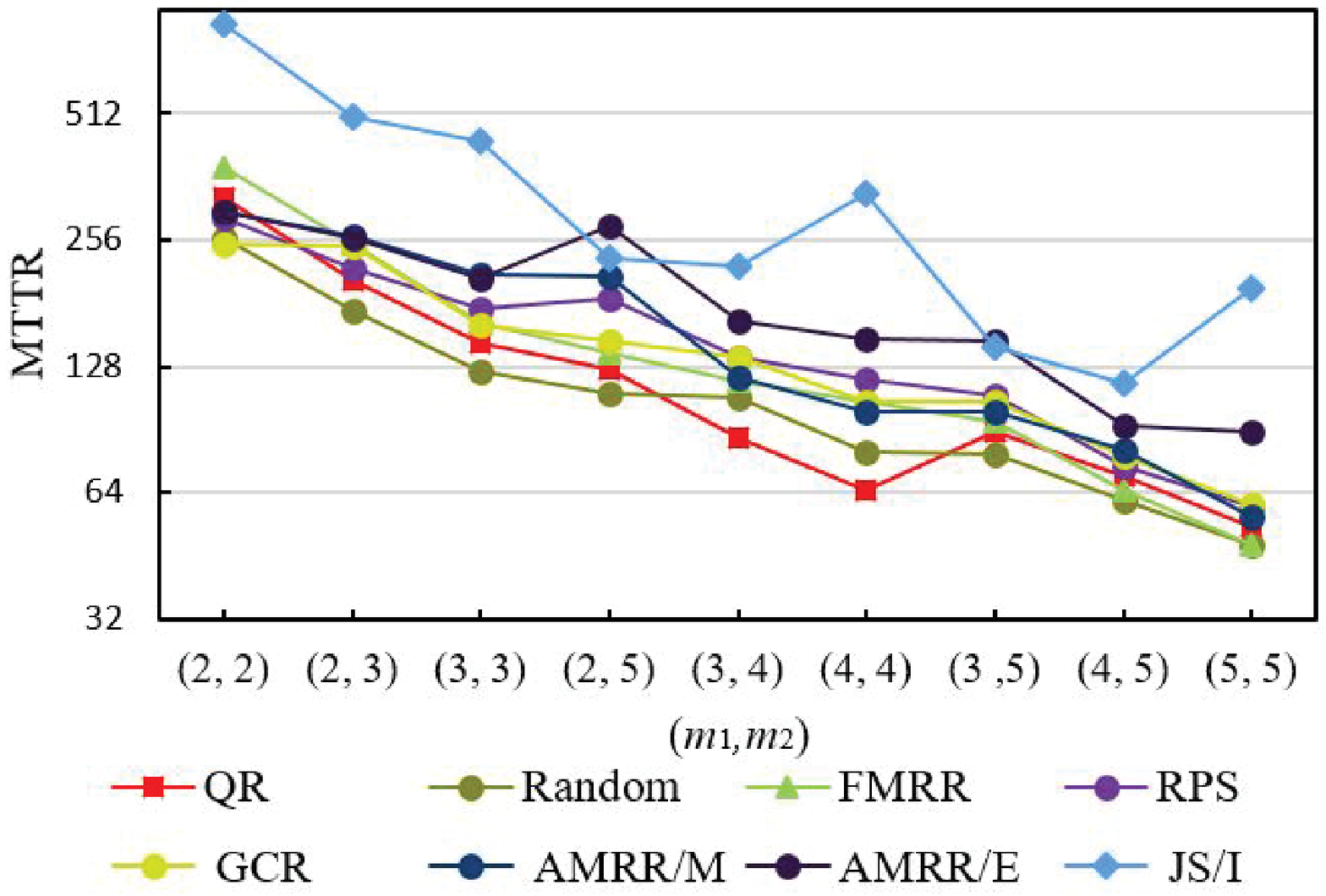}
 \\\footnotesize{ (a) MTTR for various settings of $m_1$ and $m_2$ }
    \includegraphics[width=80mm]{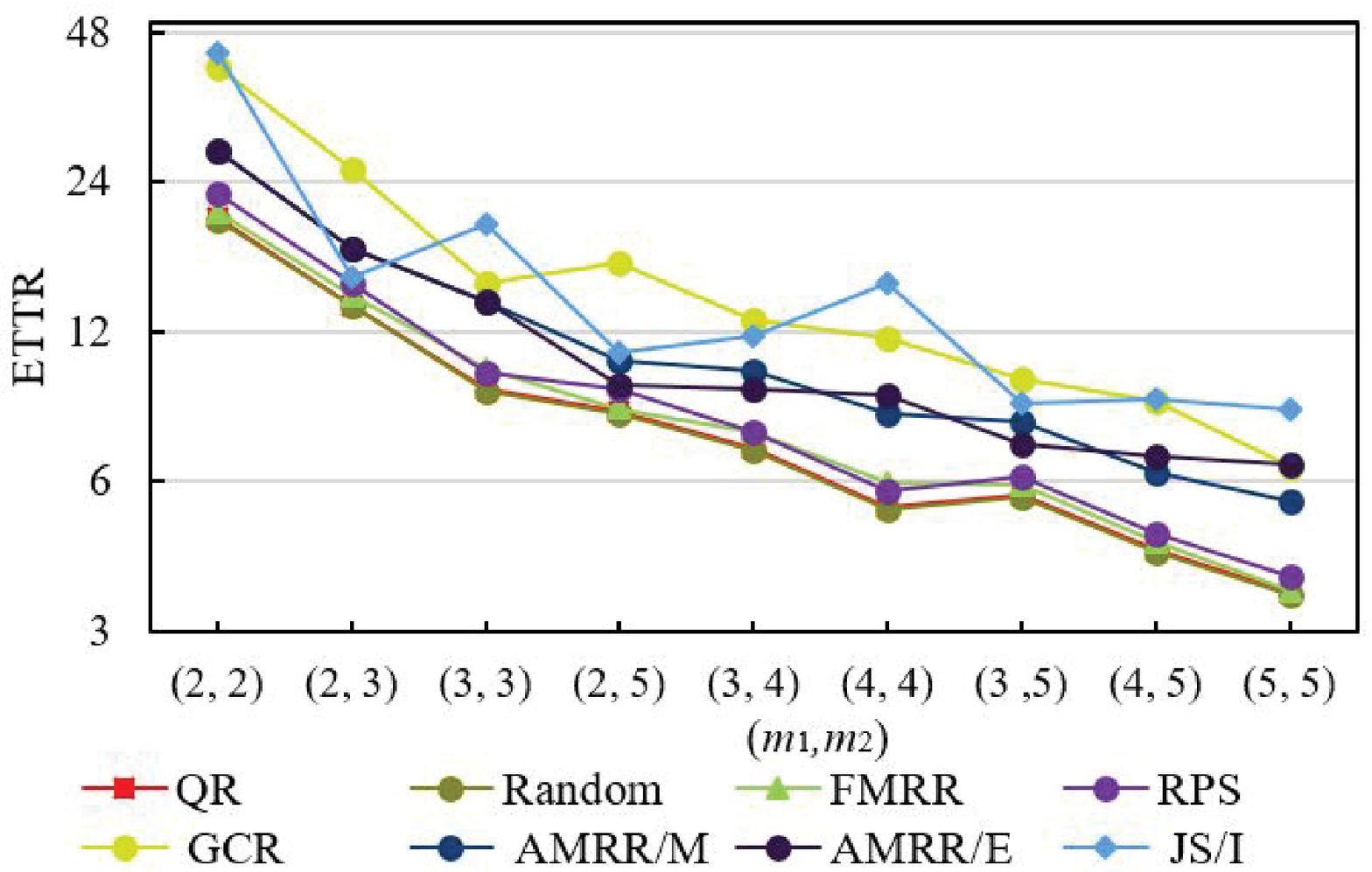}
 \\\footnotesize{ (b) ETTR for various settings of  $m_1$ and $m_2$ }
\caption{The effect of the number of radios on MTTR and ETTR for various settings of $m_1$ and $m_2$.}
    \label{fig:fig3}
\end{figure}
%\fi

%In this simulation, we fix  $N$ = 256, $n_1 = n_2$ =16 and  $G$ = 1. We then measure the MTTR and ETTR  for various settings of $(m_1,m_2)$.
%In \rfig{my6}(a), we measure the MTTR by varying the radios from (1, 1) to (8,8) for users  $SU_1$ and $SU_2$. As the number of radios increase, the MTTR and ETTR of all algorithms decreases due to availability of more channels from the increasing radios to get rendezvous.

In this simulation, we fix $N = 160$, $n_1 = n_2 = 40$, and $G = 20$. We then measure MTTR and ETTR for various settings of $(m_1, m_2)$. The simulation results are shown in \rfig{fig3}. As expected, both  MTTR and ETTR decrease when the numbers of radios $m_1$ and $m_2$ are increased. This is because the probability of finding a common channel for rendezvous  is increased when the numbers of radios $m_1$ and $m_2$ are increased. The results shown in \rfig{fig3}(a) and (b) are consistent with the findings in the simulations in the previous two settings.

%The performance of GCR, JS/P, and JS/I is the worst for almost all values of m1 or m2. Additionally, we notice that the performance of both JS/I and JS/P, degrade when m1 and m2 are the multiplicative factor of each other (ex: (2, 2), (2, 4)) as shown in \rfig{fig4}. Finally, as shown in \rfig{fig4}(b), our algorithm has the best MTTR and ETTR for the most cases in this simulation.

\bsubsec{Impact of the number of common channels}{numcommchannel}

\iffalse
% one column
%\begin{UP}
\begin{figure}[htbp]
    \centering
    \begin{tabular}{p{0.45\textwidth}p{0.45\textwidth}}
    \includegraphics[width=0.45\textwidth]{sim4a.eps} &
    \includegraphics[width=0.45\textwidth]{sim4b.eps} \\
  (a) MTTR vs. the number of common channels  & (b) ETTR vs. the number of common channels \\
    \end{tabular}
\caption{The effect of the number of common channels on MTTR and ETTR for various common channels $G$ with $n_1 = n_2$ = 64, $m_1 = m_2 = 5$.}
    \label{fig:fig4}
\end{figure}
\fi

% Two columns
%\iffalse
\begin{figure}[htbp]
    \centering
    \includegraphics[width=80mm]{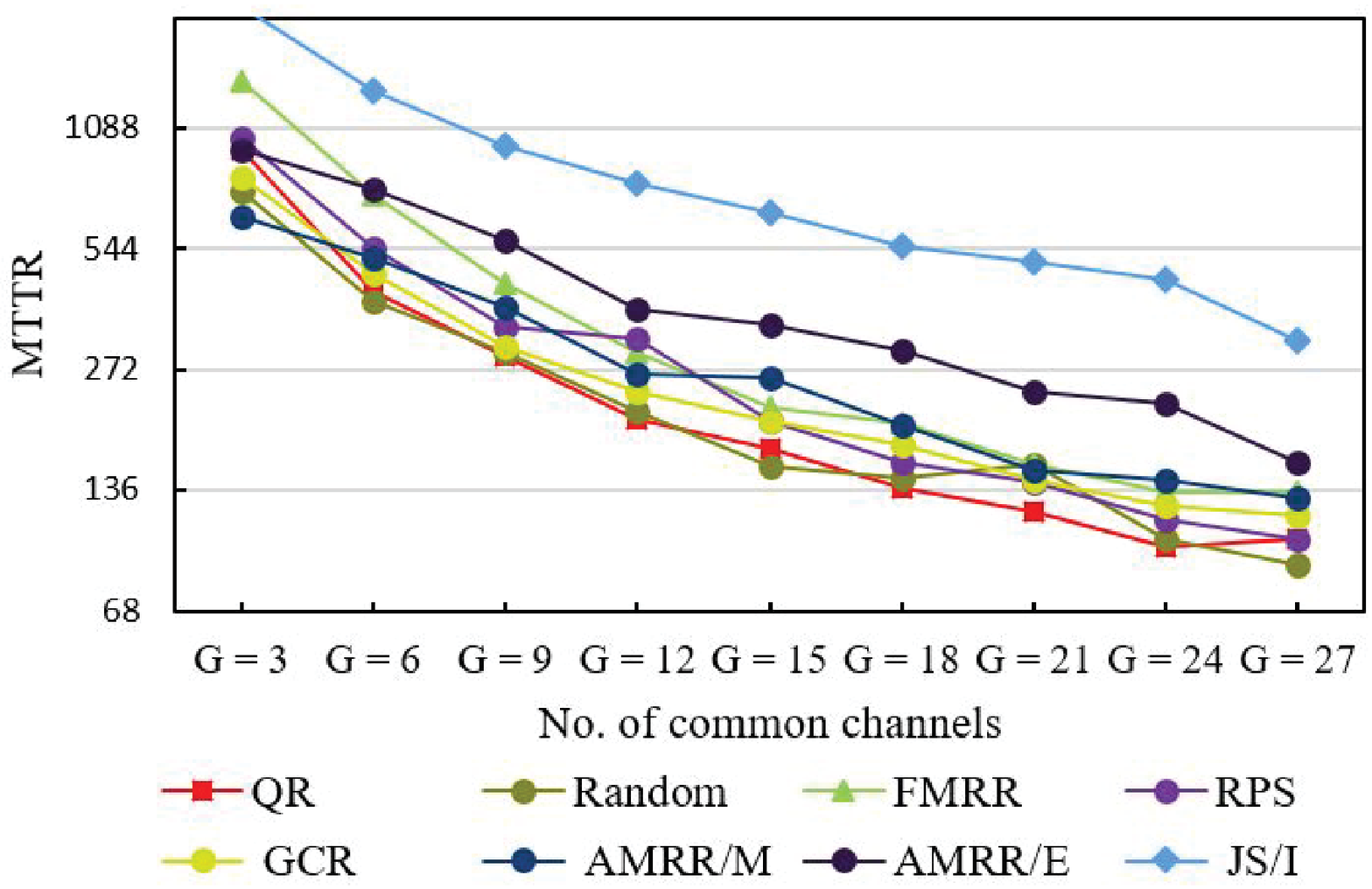}
 \\\footnotesize{ (a) MTTR vs. the number of common channels }
    \includegraphics[width=80mm]{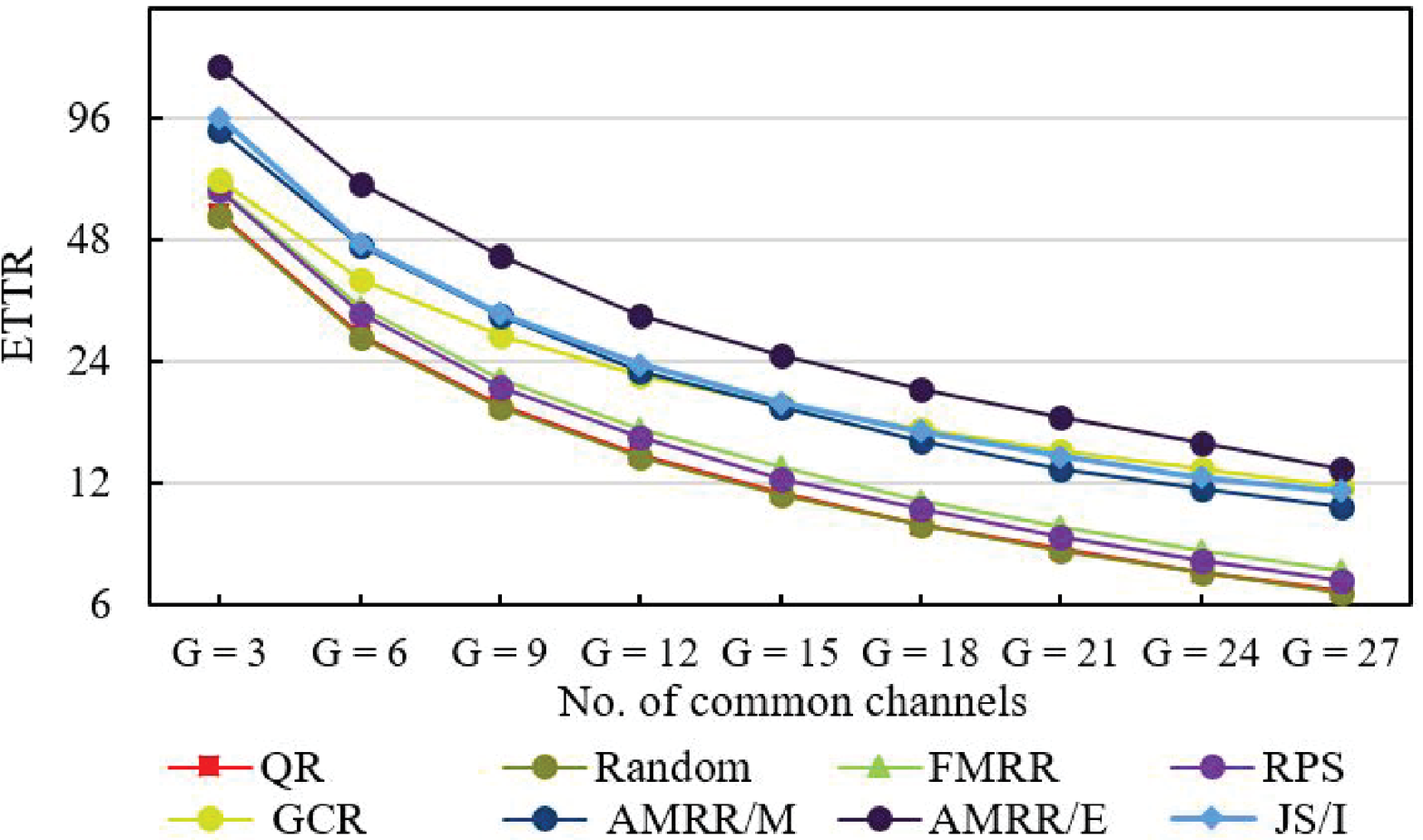}
 \\\footnotesize{ (b) ETTR vs. the number of common channels }
\caption{The effect of the number of common channels on MTTR and ETTR for various common channels $G$ with $n_1 = n_2$ = 64, $m_1 = m_2 = 5$.}
    \label{fig:fig4}
\end{figure}
%\fi

In this simulation, we fix $N = 160, n_1 = n_2 = 64 , m_1 = 5, m_2 = 5$, and vary $G$ from 3 to 27. The simulation results are shown in \rfig{fig4}.
Clearly, both MTTR and ETTR are decreasing in $G$. Once again, both the MTTR and the ETTR of our QR algorithm are almost identical to those of the random algorithm and they are better than those of the other algorithms.

% show that our MTTR has the lowest value for every instance. Similarly, as our algorithm is a quasi-random method, the ETTR stands first among all other algorithms (and is similar to the random method) that we use for comparison as shown in figure.

\bsec{Conclusion}{con}

In this paper, we proposed the quasi-random (QR) CH algorithm  that has a comparable ETTR to the random algorithm and a comparable MTTR to the best bound in the literature. Our QR algorithm does not require the unique ID assumption in \cite{Chang17} and is thus more robust to jamming attack.
It is very simple to implement in the  symmetric,  asynchronous, and heterogeneous setting with multiple radios.

There are several possible extensions of this work: (i) in this paper, we only considered using the 4B5B encoding scheme. There are other encoding schemes proposed in  \cite{Chang17} that might be also applicable to our QR algorithm. (ii) We only consider two-user rendezvous in this paper. It would be of interest to see how the QR algorithm performs in the multiuser rendezvous problem.

\begin{IEEEbiography}[{\includegraphics[width=1in,height=1.25in,clip,keepaspectratio]{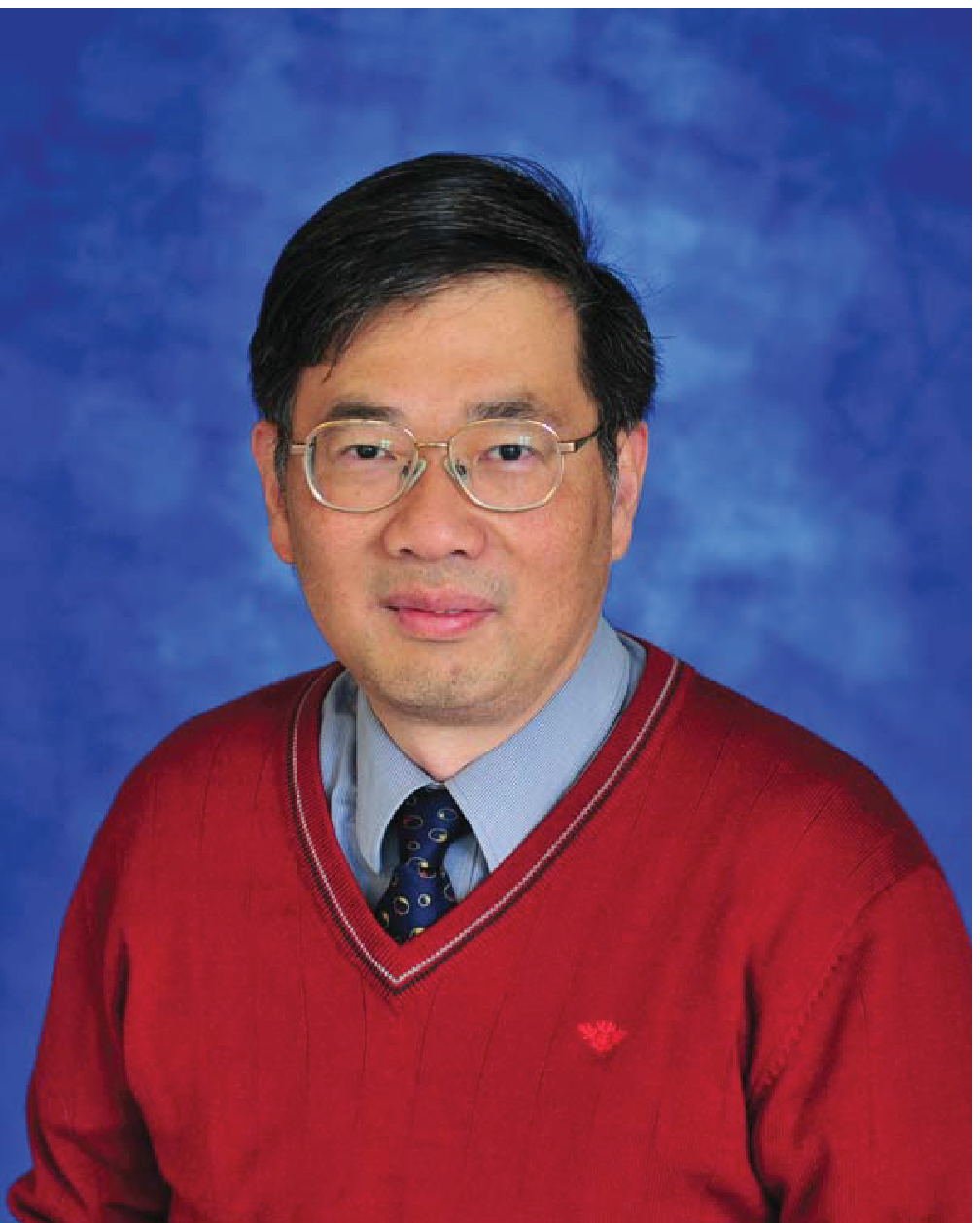}}]
{Cheng-Shang Chang} (S'85-M'86-M'89-SM'93-F'04) received the B.S. degree from National Taiwan University, Taipei, Taiwan, in 1983, and the M.S. and Ph.D. degrees from Columbia University, New York, NY, USA, in 1986 and 1989, respectively, all in electrical engineering.

From 1989 to 1993, he was employed as a Research Staff Member with the IBM Thomas J. Watson Research Center, Yorktown Heights, NY, USA. Since 1993, he has been with the Department of Electrical Engineering, National Tsing Hua University, Taiwan, where he is a Tsing Hua Distinguished Chair Professor. He is the author of the book Performance Guarantees in Communication Networks (Springer, 2000) and the coauthor of the book Principles, Architectures and Mathematical Theory of High Performance Packet Switches (Ministry of Education, R.O.C., 2006). His current research interests are concerned with network science, big data analytics, mathematical modeling of the Internet, and high-speed switching.

Dr. Chang served as an Editor for Operations Research from 1992 to 1999, an Editor for the {\em IEEE/ACM TRANSACTIONS ON NETWORKING} from 2007 to 2009, and an Editor for the {\em IEEE TRANSACTIONS ON NETWORK SCIENCE AND ENGINEERING} from 2014 to 2017. He is currently serving as an Editor-at-Large for the {\em IEEE/ACM TRANSACTIONS ON NETWORKING}. He is a member of IFIP Working Group 7.3. He received an IBM Outstanding Innovation Award in 1992, an IBM Faculty Partnership Award in 2001, and Outstanding Research Awards from the National Science Council, Taiwan, in 1998, 2000, and 2002, respectively. He also received Outstanding Teaching Awards from both the College of EECS and the university itself in 2003. He was appointed as the first Y. Z. Hsu Scientific Chair Professor in 2002. He received the Merit NSC Research Fellow Award from the National Science Council, R.O.C. in 2011. He also received the Academic Award in 2011 and the National Chair Professorship in 2017 from the Ministry of Education, R.O.C. He is the recipient of the 2017 IEEE INFOCOM Achievement Award.
\end{IEEEbiography}

\begin{IEEEbiography}[{\includegraphics[width=1in,height=1.25in,clip,keepaspectratio]{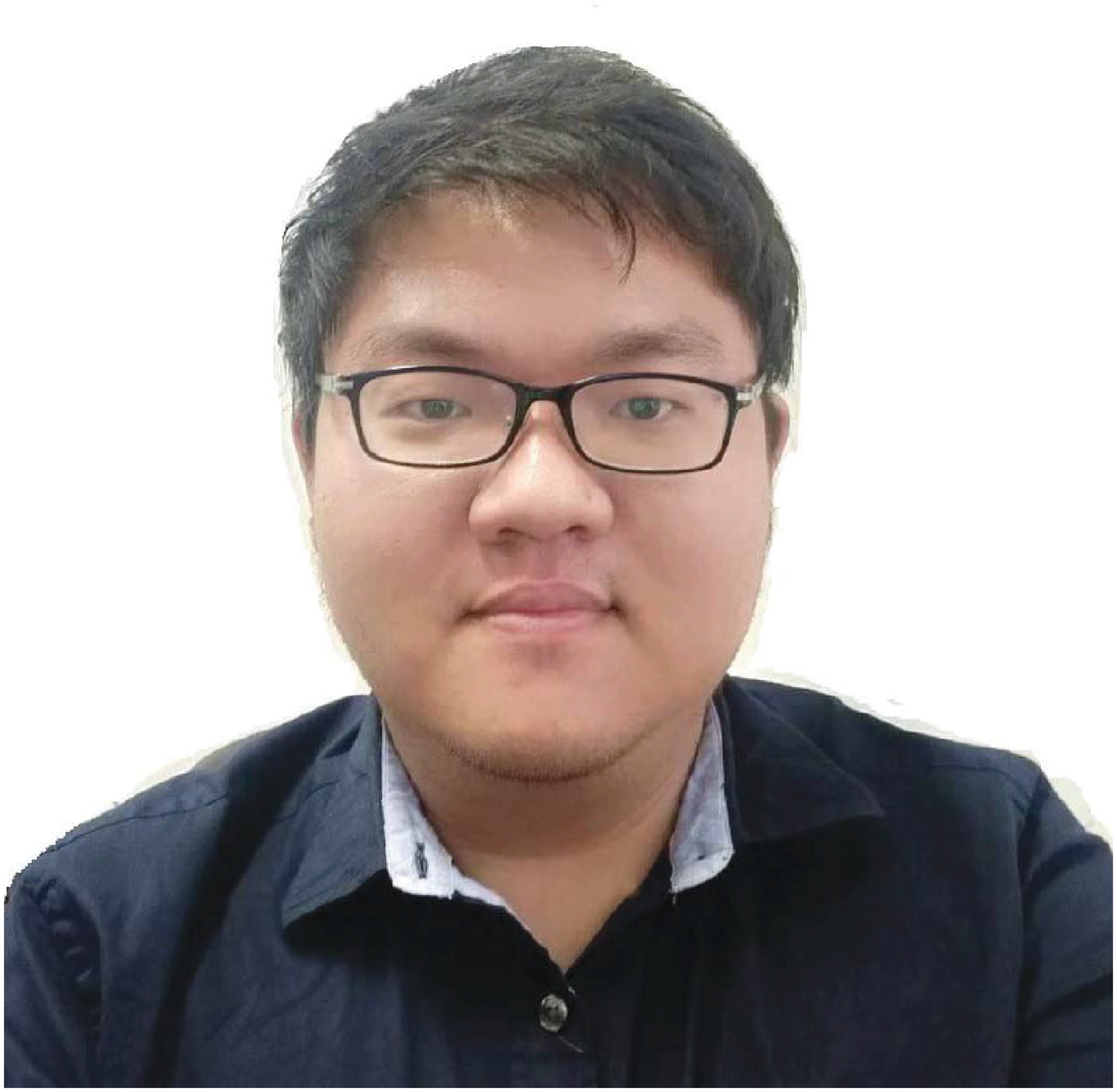}}]
{Yeh-Cheng Chang} (S'18) received B.S., degree in Computer Science from National Chung Hsing University, Taiwan. Currently he is a PhD student in the Department of Computer Science, National Tsing Hua University, Taiwan. His research interest includes cognitive radios and channel hopping sequences, wireless networks, and next generation networks.
\end{IEEEbiography}

\begin{IEEEbiography}[{\includegraphics[width=1in,height=1.25in,clip,keepaspectratio]{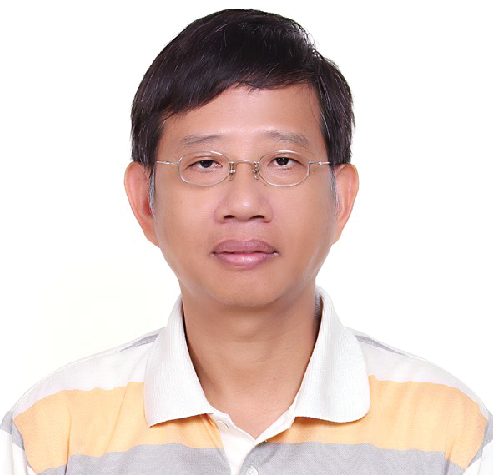}}]
{Jang-Ping Sheu}  received the B.S. degree in computer science from Tamkang University, Taiwan, Republic of China, in 1981, and the M.S. and Ph.D. degrees in computer science from National Tsing Hua University, Taiwan, Republic of China, in 1983 and 1987, respectively. He is currently a Chair Professor of the Department of Computer Science, National Tsing Hua University. He was a Chair of Department of Computer Science and Information Engineering, National Central University from 1997 to 1999. He was a Director of Computer Center, National Central University from 2003 to 2006. He was a Director of Computer and Communication Research Center from 2009 to 2015, National Tsing Hua University. He was an Associate Dean of the College of Electrical and Computer Science from 2016 to 2017, National Tsing Hua University. His current research interests include wireless communications, mobile computing, and software-defined networks. He was an associate editor of the IEEE Transactions on Parallel and Distributed Systems and International Journal of Sensor Networks. He is an Advisory Board Member of the International Journal of Ad Hoc and Ubiquitous Computing and International Journal of Vehicle Information and Communication Systems.
He received the Distinguished Research Awards of the National Science Council of the Republic of China in 1993-1994, 1995-1996, and 1997-1998. He received the Distinguished Engineering Professor Award of the Chinese Institute of Engineers in 2003. He received the K. -T. Li Research Breakthrough Award of the Institute of Information and Computing Machinery in 2007. He received the Y. Z. Hsu Scientific Chair Professor Award and Pan Wen Yuan Outstanding Research Award in 2009 and 2014, respectively. He received the Academic Award in Engineering from Ministry of Education and Medal of Honor in Information Sciences from Institute of Information and Computing Machinery in 2016 and 2017, respectively.  Dr. Sheu is an IEEE Fellow and a member of Phi Tau Phi Society.
\end{IEEEbiography}

\end{document}